\documentclass[12pt]{scrartcl}
\addtokomafont{subsection}{\large}
\addtokomafont{section}{\large}
\usepackage[utf8]{inputenc}
\usepackage[T1]{fontenc}
\usepackage[english]{babel}
\usepackage[top=25mm,bottom=25mm,left=25mm,right=25mm]{geometry}
\usepackage{lmodern}
\usepackage{microtype}
\usepackage{amsfonts}
\usepackage{amssymb}
\usepackage{amsmath}
\usepackage{amsthm}
\usepackage{color}
\usepackage{csquotes}
\usepackage{setspace}
\usepackage[backend = biber, style=authoryear,ibidtracker=context,bibencoding=utf8,hyperref=false,doi=false,url=false, maxcitenames=2, maxbibnames=50, isbn = false, firstinits = true, uniquename=false, uniquelist=false,dashed=false]{biblatex}
\renewbibmacro{in:}{}
\AtEveryBibitem{%
	\clearfield{note}%
}
\usepackage{hyperref}
\usepackage{listings}
\usepackage{xcolor}
\usepackage{textcomp}
\usepackage{graphicx}
\graphicspath{./}
\usepackage{subcaption}
\usepackage{longtable}
\usepackage{pdflscape}
\usepackage{tikz}
\usepackage{rotating}
\newtheorem{theorem}{Theorem}
\newtheorem{mydef}{Definition}

\newcommand{\changefont}[3]{
	\fontfamily{#1} \fontseries{#2} \fontshape{#3} \selectfont}
\newcommand{\E}{{\changefont{cmss}{m}{n} \operatorname{\text{E}}}}

\newcommand{\Ind}{{\changefont{cmss}{m}{n} \operatorname{\mathbf{I}}}}

\definecolor{red3}{RGB}{205,0,0}
\definecolor{blue3}{RGB}{0,0,205}
\definecolor{green3}{RGB}{0,205,0}
\definecolor{hotpink4}{RGB}{139,58,98}
\definecolor{chartreuse4}{RGB}{69,139,0}
\definecolor{darkorchid4}{RGB}{104,34,139}
\definecolor{mediumorchid4}{RGB}{122,55,139}
\definecolor{darkgreen}{RGB}{0,100,0}
\definecolor{gold4}{RGB}{139,117,0}
\definecolor{darkred}{RGB}{139,0,0}
\definecolor{iseblue}{rgb}{0,0.35,0.62}
\definecolor{darkgoldenrod4}{RGB}{139,101,8}
\definecolor{brown4}{RGB}{139,35,35}
\definecolor{goldenrod3}{RGB}{205,155,29}
\definecolor{darkorange3}{RGB}{205,102,0}

\begin{document}
\thispagestyle{empty}
\begin{center}
	
	{\LARGE{\textbf{CRIX an Index for cryptocurrencies} \footnote{Financial support from the Deutsche Forschungsgemeinschaft via CRC 649 ”Economic Risk”, IRTG 1792 ”High Dimensional Non Stationary Time Series”, as well as the Czech Science Foundation under grant no. 19-28231X, the Yushan Scholar Program and the European Union’s Horizon 2020 research and innovation program ''FIN-TECH: A Financial supervision and Technology compliance training programme'' under the grant agreement No 825215 (Topic: ICT-35-2018, Type of action: CSA), Humboldt-Universit\"at zu Berlin,  is gratefully acknowledged.}} \vspace{1cm}
		
		{\large{ Simon Trimborn \footnote{Department of Statistics \& Applied Probability, National University
					of Singapore, Singapore and Humboldt-Universit\"at zu Berlin, C.A.S.E. - Center for Applied Statistics and Economics, Spandauer
					Str. 1, 10178 Berlin, Germany, tel: +65 6516-1245, E-Mail:				simon.trimborn@nus.edu.sg}}}} \\
	{\large{ Wolfgang Karl H{\"a}rdle \footnote{Humboldt-Universit\"at zu Berlin, C.A.S.E. - Center for Applied Statistics and Economics, Spandauer
				Str. 1, 10178 Berlin, Germany and SKBI School of Business, Singapore Management University, 50
				Stamford Road, Singapore 178899, tel: +49 (0)30 2093-5630, E-Mail: haerdle@hu-berlin.de}}} \\\vspace{1cm}
	
	{\large{\today}} \\\vspace{0.8cm}
	
\end{center}

\begin{onehalfspace}
\noindent
The cryptocurrency market is unique on many levels: Very volatile, frequently changing market structure, emerging and vanishing of cryptocurrencies on a daily level. Following its development became a difficult task with the success of cryptocurrencies (CCs) other than Bitcoin. 
For fiat currency markets, the IMF offers the index SDR and, prior to the EUR, the ECU existed, which was an index representing the development of European currencies. Index providers decide on a fixed number of index constituents which will represent the market segment. It is a challenge to fix a number and develop rules for the constituents in view of the market changes. In the frequently changing CC market, this challenge is even more severe. 
A method relying on the AIC is proposed to quickly react to market changes and therefore enable us to create an index, referred to as CRIX, for the cryptocurrency market. CRIX is chosen by model selection such that it represents the market well to enable each interested party studying economic questions in this market and to invest into the market. The diversified nature of the CC market makes the inclusion of altcoins in the index product critical to improve tracking performance. We have shown that assigning optimal weights to altcoins helps to reduce the tracking errors of a CC portfolio, despite the fact that their market cap is much smaller relative to Bitcoin.
The codes used here are available via \url{www.quantlet.de} \raisebox{-1pt}{\includegraphics[scale=0.05]{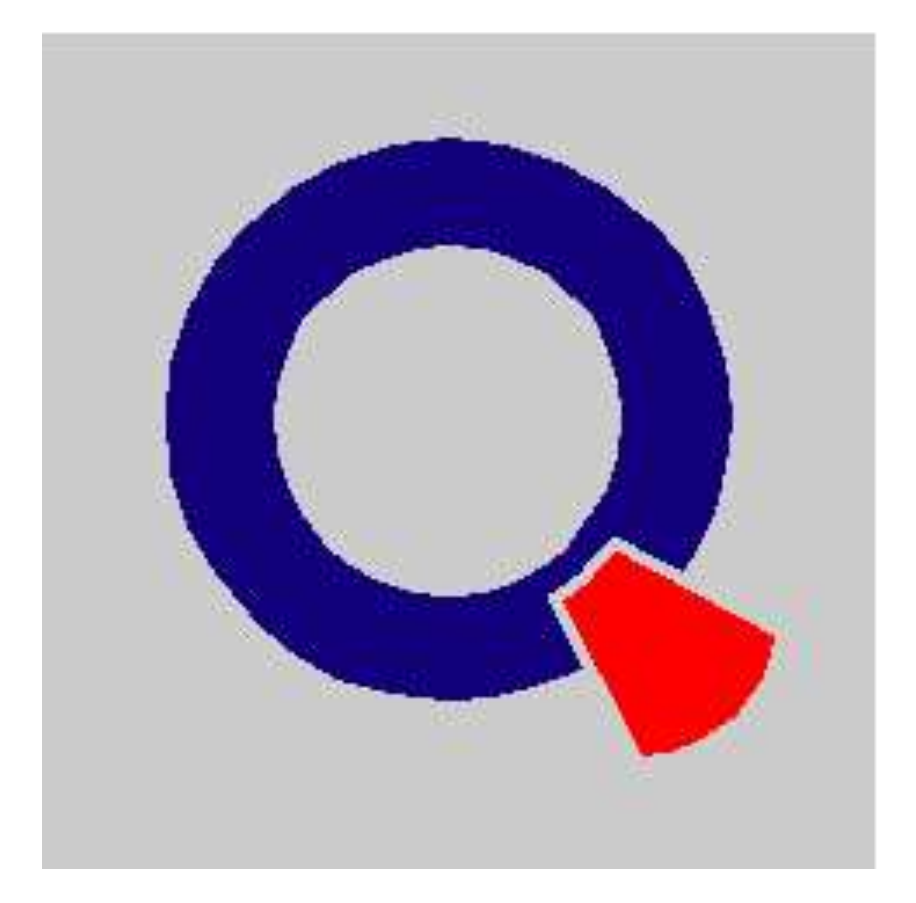}\href{www.quantlet.de}}.

\vspace{0.6cm}

\vspace{0.3cm}

This is a post-peer-review, pre-copyedit version of an article published in the Journal of Empirical Finance, Vol. 49, 107-122 (2018). The final authenticated version is available online at: \href{https://doi.org/10.1016/j.jempfin.2018.08.004}{https://doi.org/10.1016/j.jempfin.2018.08.004}

\end{onehalfspace}

\newpage
\section{Introduction}
\label{Chapter0}
More and more companies have started offering digital payment systems. Smartphones have evolved into a digital wallet, telephone companies offer banking related services: clear signal that we are about to enter the era of digital finance. In fact we are already acting inside a digital economy. The market for e-$x$ ($x$ = ``finance,'' ``money,'' ``book,'' you name it \dots) has not only picked up enormous momentum but has become standard for driving innovative activities in the global economy.  A few clicks at $y$ and payment at $z$ brings our purchase to location $w$. Own currencies for the digital market were therefore just a matter of time. Due to organizational difficulties the idea of the Nobel Laureate Hayek, see \textcite{hayek_denationalization_1990}, of letting companies offer concurrent currencies seemed for a long time scarcely feasible, but the invention of the \textit{Blockchain} has made it possible to bring his vision to life. Cryptocurrencies (CCs) have surfaced and opened up an angle towards this new level of economic interaction. Since the appearance of Bitcoins, several new CCs have spread through the Web and offered new ways of proliferation. Even states accept them as a legal payment method or part of economic interaction. E.g., the USA classifies CCs as commodities, \textcite{kawa_bitcoin_2015}, and lately Japan announced that they accept them as a legal currency, \textcite{econotimes_japans_2016}. Obviously, the crypto market is fanning out and shows clear signs of acceptance and deepening liquidity, so that a closer look at its general moves and dynamics is called for. 

The transaction graph of Bitcoin (BTC), the Blockchain, has received much attention, see e.g. \textcite{ron_quantitative_2013} and \textcite{reid_analysis_2013}. Even the economics of BTC has been studied, e.g. \textcite{bolt_value_2016} and \textcite{kristoufek_what_2015}. To our best knowledge, the development of the entire CC market has not been studied so far, only subsamples have been taken into account. \textcite{wang_buzz_2017} studied the variations of 5 CCs. \textcite{elendner_cross-section_2017} analyzed the top 10 CCs by market capitalization and found that their returns are weakly correlated with each other. Furthermore, a Principal Component (PC) Analysis, carried out in the same reference, showed 7 out of 10 PC were necessary to describe more than 90\% of the variance. These findings indicate the price evolution of CCs is very different from each other. This brings us to the conclusion that BTC, even though it dominates the market in terms of its market capitalization, can not lead the direction of the market. The movements of other CCs are important too, when one analyzes the market. 
Having a closer look at the different CCs, it becomes obvious they have different kind of missions and technical aspects. Bitcoin pioneered as the token of the first decentralised, distributed ledger,  giving start to multiple interpretations of its nature and purpose: new type of currency, commodity (like gold), alternative asset or innovative technology. The currently second most important CC by market capitalization - Ethereum - was  created with a particular goal in mind - to power the blockchain based Ethereum platform for  company building (DAO) and smart contract implementation. This idea triggered an unprecedented interest as it allowed companies to enter the field without creating their own blockchain ecosystem. Newcomers could benefit from the existing supporters of the respective platform,  which allowed faster entry, adoption and operation. Other CCs, like Ripple (XRP),  are intended to fuel the transaction network bridging traditional markets (banks) and the crypto ecosystem. Ripple also became one of the first successful cases of pre-emitted CC, abandoning the idea of decentralisation. Since the appearance of BTC many technological advancements took place. Some CCs are designed for faster (or even immediate) transactions, like Litecoin (LTC), some are more efficient energy-wise, like DASH. Many embraced different hashing algorithms, altering the mining process, like Monero. Long ASIC domination is being disrupted, Proof-of-work is replaced by Proof-of-Stake, new ways to motivate those providing computational power are introduced. Regardless the type of CC, one witnesses a new kind of transaction network with a different approach for fees and handling of trust issues. The intended and actual usage can be interpreted as the business model of the different CCs and the participation in  either CC can give advantages over others, \textcite{white_market_2014}. 

In the first month of 2017, CCs other than BTC (altcoins) showed a strong gain in their market capitalization, reducing the dominance of BTC in the market. The finding of very different movements of CCs and the stronger position of alternative CCs in the market infers the necessity of a market index for the CC market for tracking the market movements. Comparing CCs against a market index answers economic questions like which business model is more successful than another one, gained recently compared to other CCs, drives the success of the market, is more established.
Comparing a CC market index against other market indices answers economic and financial questions like which market proxy is more volatile, has more tail risk, attracts more investments.
We construct CRIX, a market index (benchmark) which will enable each interested party to study the outlined economic questions, the performance of the CC market as a whole or of single CCs. Studying the stochastic dynamics of CRIX will allow a la limite to create ETFs or contingent claims.

Many index providers construct their indices with a fixed number of constituents, see e.g. \textcite{ftse_ftse_2016}, \textcite{s&p_index_2014} and \textcite{deutsche_boerse_ag_guide_2013}. If the respective index is intended to be a proxy for the performance of a market, this requires huge trust from economists and investors into the choice of the index constituents by the index provider. On the other hand, the CRSP index family, derived for the US market, \textcite{crsp_crsp_2015}, has no boundary on the number of index constituents. The number of constituents is reviewed daily and adjusted until the index members cover a predefined share of the market capitalization. Such a dynamic methodology is important in the market of CCs since the number of CCs changes daily. Additionally the market value of CCs often changes frequently, which increases the market volatility and therefore the need for considering such a CC for the representation of the market. Our intention is extending the idea behind the CRSP indices. Our first goal is constructing a methodology for CRIX which relies on model selection criteria to receive a proxy for the market and to replace the trust problematic with a statistical methodology. The resulting methodology is dynamic in the number of index constituents, like the CRSP indices. By this method only CCs which add informative value to the index are considered, which makes it representative. If more CCs than BTC are necessary to fulfill this requirement, they will be added. However we are concerned with the dominance of BTC in an index solely relying on market capitalization. Thus we introduce a second weighting scheme based on weighting by trading volume. Due to the usage of trading volume, the respective index is constructed in terms of trading focus. If the market participants focus more on altcoins than on BTC, these receive a higher weight. On the other hand, if the market focus is truly on BTC, it will receive a high weight in either index. Our second goal, constructing an investable index will be fulfilled by the methodology itself due to having a sparse index, only consisting of actively traded CCs in a market with low transaction costs. Note that due to the low transaction costs in the CC market, a dynamic methodology creates low additional costs. Additionally to the methodology ensuring an investable index, the proposed trading volume weighting scheme further supports this goal.  

Investing into an ETF composed of the constituents of CRIX implies some differences compared to traditional index investing. In the traditional setting only the constituents are reviewed and replaced on the review date - if necessary - according to the index rules. In dynamic index investing the constituents are also reviewed for their number. This requires the manager of the fund to buy and sell more assets on the review date. In a market with high transaction costs, this approach is more costly. But the market of CCs has very low transaction costs, thus this problem won't occur in this market.

To compute CRIX, the differences in the log returns of the market against a selection of possible indices is evaluated. The results show, that the AIC works well to evaluate the differences. It penalizes the index for the number of constituents. %, so definitions \ref{def_benchmark} and \ref{def_portfolio} are met. 
For the calculation of the respective likelihoods, a non-parametric approach using the \textcite{epanechnikov_non-parametric_1969} kernel is applied. The proof for the impact of the value of an asset in the market on the AIC method is given, thus a top-down approach is applied to select the assets for the benchmarks to choose from, where the sorting depends on either market cap or trading volume. The number of constituents is recalculated quarterly to ensure an up-to-date fit to the current market situation. With CRIX one may study the contingent claims and the stochastic nature of this index, \textcite{chen_econometric_2017}, or study the CC market characteristics against traditional markets, \textcite{hardle_crix_2015}.

This paper is structured as follows. Section \ref{Chapter1} introduces the topic and reviews the basics of index construction. In Section \ref{Chapter2} the method for dynamic index construction for CRIX is described and Section \ref{Chapter3} introduces the remaining rules for CRIX. Section \ref{Chapter4} describes further variants to create a CRIX family. Their performance is tested in Section \ref{Chapter5}. In Sections \ref{Chapter6} and \ref{Chapter6.1} the new method is applied to the German and Mexican stock markets to check the performance of the methodology against existing indices. The codes used to obtain the results in this paper are available via \url{www.quantlet.de} \raisebox{-1pt}{\includegraphics[scale=0.05]{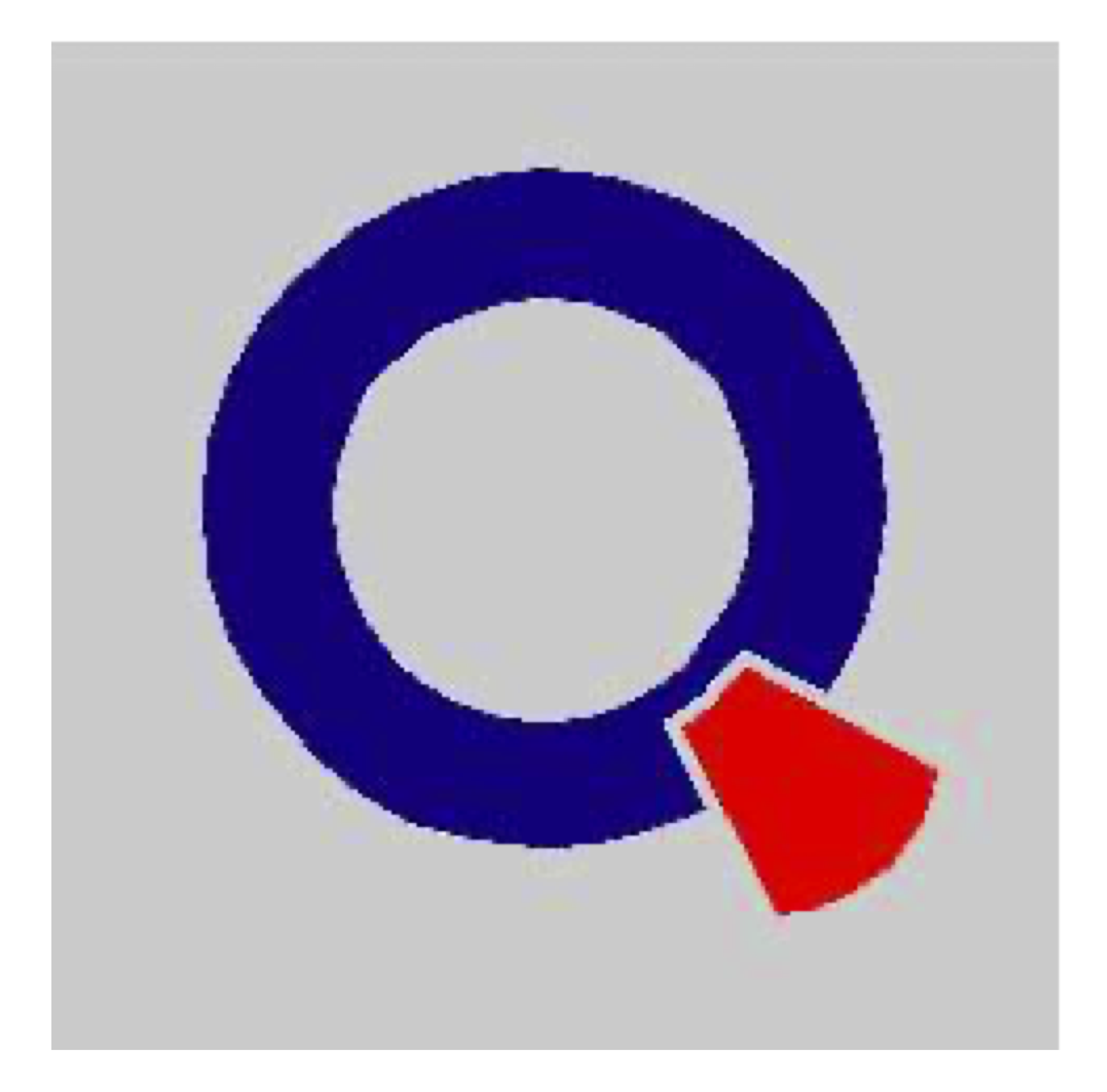}\href{www.quantlet.de}}.

\section{Index construction}
\label{Chapter1}
\label{chap_CRIX_construction}
\begin{figure}
	\begin{center}
		\includegraphics[scale = 0.5]{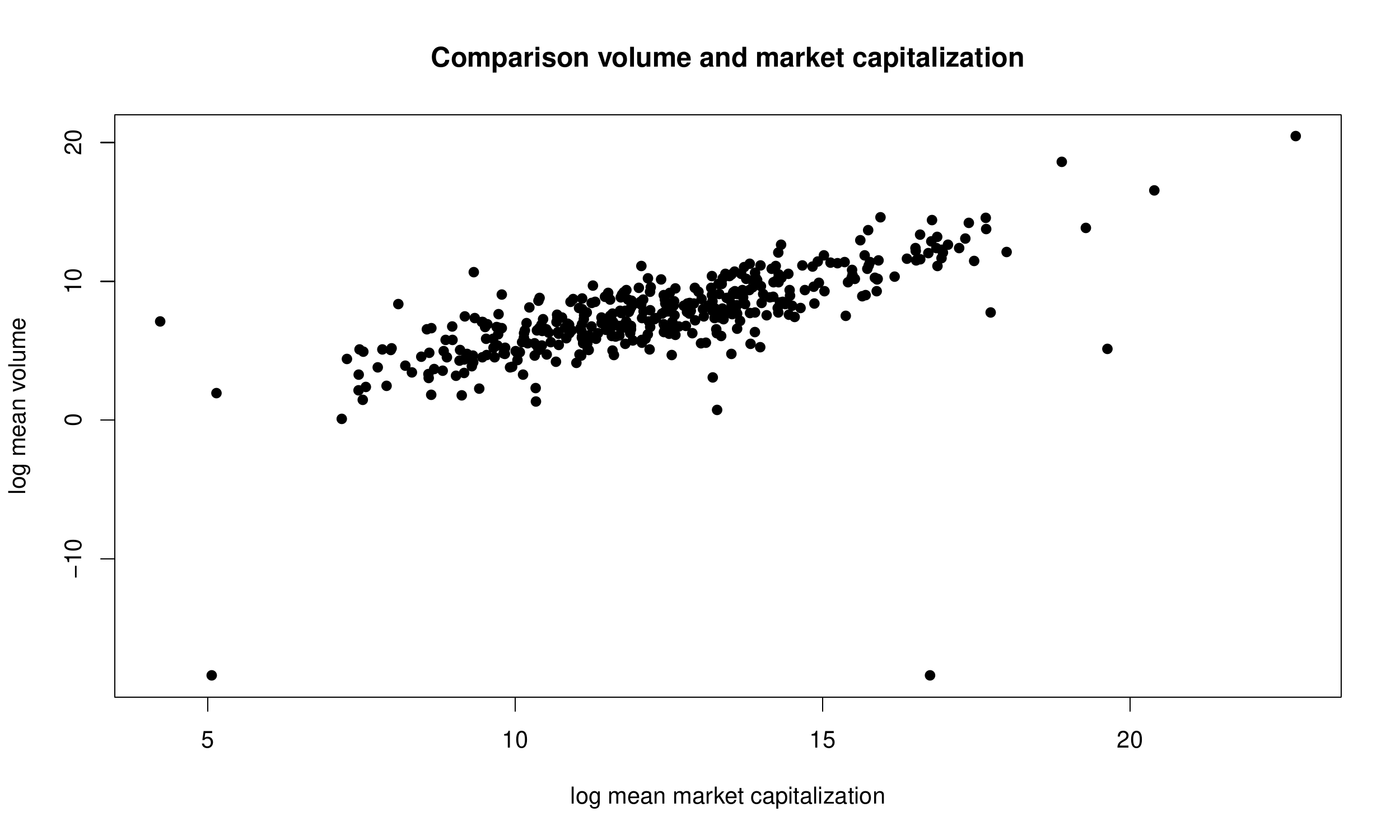}
		\caption{Comparison of the log mean trading volume and log mean market capitalization, both measured in USD, for all CCs in the dataset over the time period 20140401 - 20170325 }\hfill\raisebox{-1pt}{\includegraphics[scale=0.05]{quantlet} \href{https://github.com/QuantLet/CRIX}{VolMarketCapComparison}}
		\label{plot_VolMarketCapComparison}
	\end{center}
\end{figure}

The basic idea of any price index is to weight the prices of its constituent goods by the quantities of the goods purchased or consumed. The Laspeyres index takes the value of a basket of $k$ assets and compares it against a base period:
\begin{equation}
\label{formula_laspeyres}
P_{0t}^L(k) = \frac{\sum_{i=1}^k P_{it} Q_{i0}}{\sum_{i=1}^k P_{i0} Q_{i0}}
\end{equation}
with $P_{it}$ the price of asset $i$ at time $t$ and $Q_{i0}$ the quantity of asset $i$ at time $0$ (the base period). 
For market indices, such as CRSP, S\&P500 or DAX, the quantity $Q_{i0}$ is the number of shares of the asset $i$ in the base period. Multiplied with its corresponding price, the market capitalization results, hence the constituents of the index are weighted by their market capitalizations. These indices are often referred to as benchmarks for their respective market. We define the term benchmark: 
\begin{mydef}
	\label{def_benchmark}
	A benchmark is a measure which consists of a selection of CCs that are representing the market.
\end{mydef}
But markets change. A company which was representative for market developments yesterday might no longer be important today. On top of that, companies can go bankrupt, a corporation can raise the number of its outstanding shares, or trading in it can become infrequent. All these situations must produce a change in the index structure, so that the market is still adequately represented. Hence companies have to drop out of the index and have to be replaced by others. The index rules determine in which cases such an event happens. The formula of Laspeyres (\ref{formula_laspeyres}) can not handle such events entirely because a change of constituents will result in a change in the index value that is not due to price changes. Therefore, established price indices like DAX or S\&P500, see \textcite{deutsche_boerse_ag_guide_2013} and \textcite{s&p_index_2014} respectively, and the newly founded index CRIX$(k)$, a CRyptocurrency IndeX, \href{http://thecrix.de/}{thecrix.de}, use the adjusted formula of Laspeyres,
\begin{equation} 
\label{formula_CRIX}
\text{CRIX}_t(k,\beta) = \frac{\sum_{i=1}^k \beta_{i,t^-_l} P_{it} Q_{i,t^-_l}}{Divisor(k)_{t^-_l}}
\end{equation}
with $P$, $Q$ and $i$ defined as before, $\beta_{i,t^-_l}$ the adjustment factor of asset $i$ found at time point $t_l^-$, $l$ indicates that this is the $l$-th adjustment factor, and $t^-_l$ the last time point when $Q_{i,t^-_l}$, $Divisor(k)_{i,t^-_l}$ and $\beta_{i,t^-_l}$ were updated. In the classical setting, $\beta_{i,t^-_l}$ is defined to be $\beta_{i,t^-_l} = 1$ for all $i$ and $l$. Anyhow, some indices use $\beta_{i,t^-_l}$ to achieve maximum weighting rules, e.g. \textcite{deutsche_boerse_ag_guide_2013} and \textcite{mexbol_prices_2013}. The $Divisor$ ensures that the index value of CRIX has a predefined value on the starting date. It is defined as
\begin{equation}
Divisor(k,\beta)_0 = \frac{\sum_{i=1}^k \beta_{i0} P_{i0} Q_{i0}}{\text{starting value}}.
\end{equation}
The starting value could be any possible number, commonly 100, 1000 or 10000. It ensures that a positive or negative development from the base period will be revealed.
Whenever changes to the structure of CRIX occur, the $Divisor$ is adjusted in such a way that only price changes are reflected by the index. Defining $k_1$ and $k_2$ as number of constituents, it results
\begin{equation}
\label{formula_CRIXDivisor}
\frac{\sum_{i=1}^{k_1} \beta_{i,t^-_{l-1}} P_{i, t-1}Q_{i, t^-_{l-1}}}{Divisor(k_1,\beta)_{t^-_{l-1}}} = \text{CRIX}_{t-1}(k_1,\beta) = \text{CRIX}_t(k_2,\beta) = \frac{\sum_{j=1}^{k_2} \beta_{j,t^-_l} P_{j, t}Q_{j, t^-_l}}{Divisor(k_2,\beta)_{t^-_l}}.
\end{equation}
In indices like FTSE, S\&P500 or DAX the number of index members is fixed, $k_1 = k_2$, see \textcite{ftse_ftse_2016}, \textcite{s&p_index_2014} and \textcite{deutsche_boerse_ag_guide_2013}. As long as the goal behind these indices is the reflection of the price development of the selected assets, this is a straightforward approach. But, e.g., DAX is also meant to be an indicator for the development of the market as a whole, see \textcite{jansen_deutsche_1992}. This raises automatically the question of whether the included assets and the weighting scheme are representing the market. Since the constituents are chosen using a top-down approach, meaning that the biggest companies by market capitalization are included, the intuitive answer is yes. But it leaves a sour taste that additional assets may describe the market more appropriately. Furthermore different weighting schemes provide another view on the market. One may object by referring to total market indices like the Wilshire 5000, S\&P Total Market Index or CRSP U.S. Total Market Index, see \textcite{wilshire_associates_wilshire_2015}, \textcite{s&p_dow_2015} and \textcite{crsp_crsp_2015}, that are providing a full description. But financial practice has shown that smaller indices like DAX30 and S\&P500 receive more attention in evaluating the movements of their corresponding markets, probably because they are easier to invest in due to the smaller number of constituents.
It is therefore appealing to know which are the representative assets in a market and which smaller number of index constituents eases the handling of a tracking portfolio. Additionally, one may be concerned that an index would include illiquid and non-investable assets which makes the management of a tracking portfolio even more difficult. Figure \ref{plot_VolMarketCapComparison} shows that this is indeed a problem in the CC market. Some CCs have a fairly high market capitalization while their respective trading volume is very low. This is problematic, because an asset which is not frequently traded can not add enough information to a market index to display market changes and is difficult to trade for an investor. Hence, one goal behind constructing CRIX is making it investable by concentrating on liquid CCs:
\begin{mydef}
	\label{def_portfolio}
	Between investment portfolios with equal performance, the one with the least assets is preferable.
\end{mydef}

We react to the goals and problems in two ways: First, these thoughts raise the question which value of $k$ is "optimal" for building an investable benchmark for the market. Additionally, especially young and innovative markets may change their structure over time. 
Therefore, a quantification of an accurate CC benchmark with sparse number of constituents is asked for.
Since the CC market shows a frequently changing market structure with a huge number of illiquid CCs, a time varying index selection structure is applied. The later described selection method omits illiquid CCs by construction, because only CCs who show changes in their return series can be selected to be added to CRIX by the method. Due to the low transaction costs in this market, a dynamic methodology is applicable since it does not raise the costs of restructuring a tracking portfolio too much. Secondly, we apply two kind of weighting schemes, Table \ref{weighting}. We apply the classical setting to build a proper market index which is only flexible in terms of the dynamic constituents and tackles the illiquidity issue due to the applied selection method. The liquidity weighting allows one to weight CCs higher, which are more traded relative to their market capitalization and therefore implicitly acquire more financial attention. This weighting scheme bails (\ref{formula_CRIX}) down to weighting the price development by their trading volume, 
\begin{equation}
\label{formula_LCRIX}
\text{LCRIX}_t(k,\beta) = \frac{\sum_{i=1}^k \frac{Vol_{i, t^-_{l}} }{ P_{i, t^-_{l}}Q_{i, t^-_{l}} } P_{it} Q_{i,t^-_l}}{Divisor(k)_{t^-_l}} = \frac{\sum_{i=1}^k \frac{Vol_{i, t^-_{l}} }{ P_{i, t^-_{l}} } P_{it}}{Divisor(k)_{t^-_l}} .
\end{equation}
The latter is referred to as Liquidity CRIX (LCRIX). This approach has the potential to diminish the influence of e.g. Bitcoin stronger than the market cap weighting, if the relation of trading volume to market cap is higher for other CCs. In section \ref{Chapter5} we show that LCRIX has a better mean directional accuracy than CRIX and puts more weight on altcoins, Table \ref{table_weights}, therefore tackling the issue of BTC dominance when the actual trading amount suggests a different result.

\begin{table}[ht]
	\centering
	\begin{tabular}{r|cc}
		\hline \hline
		& market cap weighting & liquidity weighting \\
		\hline
		$\beta_{i,t^-_l}$ & 1 & $\frac{Vol_{i, t^-_{l}} }{ P_{i, t^-_{l}}Q_{i, t^-_{l}} }$ \\
		\hline \hline
	\end{tabular}
\caption{Weighting schemes for derivation of CRIX}
\label{weighting}
\end{table}

\section{Dynamic index construction}
\label{Chapter2}
This section is dedicated to describing the composition rule which is used to find the number of index members---the spine of CRIX and LCRIX. 
Since CRIX will be a benchmark for the CC market, the dimension and evaluation of the market has to be defined:
\begin{mydef}
	\label{def_total_market}
	The total market (TM) consists of all CCs in the CC universe. Its value is the combined market value of the CCs.
\end{mydef}
To compare the TM with a benchmark candidate, it will be normalized by a Divisor,
\begin{equation} \label{formula_total_market}
	\text{TM}(K)_t = \frac{\sum_{i=1}^K P_{it} Q_{i,t^-_l}}{Divisor(K)_{t^-_l}}
\end{equation}
with $K$ the number of all CCs in the CC universe. Note that no adjustment factor is used for TM$(K)_t$. For the volume weighting, the TM is defined as LTM respectively, 
\begin{equation} \label{formula_total_market_volume}
\text{LTM}(K)_t = \frac{\sum_{i=1}^K \frac{Vol_{i, t^-_{l}} }{ P_{i, t^-_{l}}Q_{i, t^-_{l}} } P_{it} Q_{i,t^-_l}}{Divisor(K)_{t^-_l}}. 
\end{equation}
In the further explanations, the focus lies on the TM. However when LCRIX is derived, it is optimized against LTM. The results can be easily extended to the case of LTM.
Further define the log returns:

\begin{align}
\varepsilon(K)_{t}^{TM} &= \log \{\text{TM}(K)_t\} - \log \{\text{TM}(K)_{t-1}\} \\
\varepsilon(k,\beta)_{t}^{CRIX} &= \log \{\text{CRIX}(k,\beta)_t\} - \log \{\text{CRIX}(k,\beta)_{t-1}\},
\end{align}
where $\text{CRIX}(k,\beta)_{t}$ is the CRIX with $k$ constituents at time point $t$.

The goal is to optimize $k$ and $\beta$ so that a sparse but accurate approximation in terms of 
\begin{align}
\label{formula_min_eps}
\underset{k,\beta}{\min} \,\, \| \varepsilon(k,\beta) \|^2  = \underset{k,\beta}{\min} \,\, \| \varepsilon(K)^{TM} - \varepsilon(k,\beta)^{CRIX} \|^2, 
\end{align}
is achieved, where $\varepsilon(k,\beta)$ is the difference in the log returns of TM($K$) and CRIX($k,\beta$). A squared loss function is chosen in (\ref{formula_min_eps}), since it heavily penalizes deviations. 

Since the value of $\text{TM}(K)_t$ is unknown and not measurable due to a lack of information, the total market index will be defined and used as a proxy for the TM$(K)$. The definition is inspired by total market indices like \textcite{crsp_crsp_2015}, \textcite{s&p_dow_2015} and \textcite{wilshire_associates_wilshire_2015}. They use all stocks for which prices are available.

\begin{mydef}
	\label{def_TotalMarketIndex}
	The total market index (TMI) contains all CCs in the CC universe for which prices are available. The CCs are weighted by their market capitalization.
\end{mydef}
This changes (\ref{formula_total_market}) to
\begin{equation*}
\text{TMI}_t(k_{max}) = \frac{\sum_{i=1}^{k_{max}} P_{it} Q_{i,t^-_l}}{Divisor(k_{max})_{t^-_l}}
\end{equation*}
with $k_{max}$ the maximum number of CCs with available prices and (\ref{formula_min_eps}) to
\begin{align}
	\label{formula_min_eps_hat}
	\underset{k,\beta}{\min} \,\, \| \widehat{\varepsilon}(k,\beta) \|^2  = \underset{k,\beta}{\min} \,\, \| \varepsilon(k_{max})^{TMI} - \varepsilon(k,\beta)^{CRIX} \|^2 \\
	\text{s.t.: } 1 \leq k \leq k^u \nonumber \\
	k = k_1 + s \\
	k^u \in [1,k_{max}] \nonumber \\
	s \in [1,k_{max}-k_1] \nonumber \\
	\beta^{1\times k} = (1,\dots,1,\beta_{k_1+1},\dots,\beta_{k_1+s})^\top \nonumber \\
	\beta_{k_1+1}, \dots, \beta_{k_1+s} \in (-\infty, \infty), \nonumber
\end{align}

where $\varepsilon(k_{max})^{TMI}$ are the log returns for TMI. In the derivation of LCRIX, the optimization is performed against LTMI and $\beta^{1\times k} = (\beta_1,\dots,\beta_k,\beta_{k_1+1},\dots,\beta_{k_1+s})^\top$ where $\beta_i = \frac{Vol_{i, t^-_{l}} }{ P_{i, t^-_{l}}Q_{i, t^-_{l}} }$ for $i=1,\dots,k_1$ and $\beta_{k_1+1}, \dots, \beta_{k_1+s} \in (-\infty, \infty)$.

Several constraints were introduced with (\ref{formula_min_eps_hat}). The parameters $\beta_{k+1}, \dots, \beta_{k+s}$ are included to evaluate if adding $s$ more assets to the index explains the difference between $\varepsilon(k_{max})^{TMI}$ and $\varepsilon(k,\beta)^{CRIX}$ better. The first $k$ assets ($k_1$) won't be adjusted by a parameter, so no parameter estimation is necessary. This makes the first term a constant. The choice of $k_1$ is important since it defines the number of base CCs to be included in the index. The parameters of the next $s$ assets have to be estimated, so (\ref{formula_CRIX}) becomes
\begin{equation*}
\text{CRIX}_t(k,\beta) = \frac{\sum_{i=1}^{k_1} P_{it} Q_{i,t^-_l} + \sum_{j=k_1+1}^{k_1+s} \beta_{j,t^-_l} P_{jt} Q_{j,t^-_l}}{Divisor(k_1)_{t^-_l}}.
\end{equation*}

A number of criteria are applicable. Model selection (SC) criteria can be categorized by their property to be either asymptotic optimal or consistent in choosing the true model. In this context will be investigated: Generalized Cross Validation (GC), Generalized Full Cross Validation (GFC), Mallows' C$_p$, Shibata (SH), Final Prediction Error (FPE) and Akaike Information Criterion (AIC), all asymptotic optimal criteria under the assumption of Gaussian distributed residuals. Since CRIX is supposed to be a benchmark model, all possible models under certain restrictions for the number of parameters are included in the test set, 
\begin{equation}
\label{equation_Theta_SC}
\Theta_{SC} = \{\text{CRIX}(k_1,\beta), \text{CRIX}(k_2,\beta), \dots \},
\end{equation}
where $k_1, k_2, \dots$ are predefined values and $SC \in \{\text{GC}, \text{GFC}, \text{C}_p, \text{SH}, \text{FPE}, \text{AIC}\}$. Recall that the intention behind CRIX is to discover under a squared loss function the best model to describe the data (benchmark), which supports the choice of an asymptotic optimal criteria. 
The GC criterion, see \textcite{craven_smoothing_1978}, is defined as 
\begin{equation}
\text{GC}\{\widehat{\varepsilon}(k,\beta),s\} = \frac{ T^{-1} \sum_{t=1}^T \widehat{\varepsilon}(k,\beta)_t^2 }{ (1 - T^{-1}s)^2 }
\end{equation}
by assuming that $s < T$. One shall note that $s$ and not $k+s$ defines the number of variables to penalize for, since $k$ parameters are set to be 1 and need not be estimated. According to \textcite{arlot_survey_2010}, the asymptotic optimality of GC was shown in several frameworks. The GFC, see \textcite{droge_comments_1996}:
\begin{equation}
\text{GFC}\{\widehat{\varepsilon}(k,\beta),s\} = T^{-1} \sum_{t=1}^T \widehat{\varepsilon}(k,\beta)_t^2 (1 + T^{-1}s)^2
\end{equation}
is an alteration.

A further score, SH,
\begin{equation}
\text{SH}\{\widehat{\varepsilon}(k,\beta),s\} = \frac{T + 2s}{T^2}\sum_{t=1}^T \widehat{\varepsilon}(k,\beta)_t^2,
\end{equation}
was shown to be asymptotically optimal, \textcite{shibata_optimal_1981}, and asymptotically equivalent to Mallows' C$_p$ and AIC.

\textcite{mallows_comments_1973}' C$_p$:
\begin{equation}
	\text{C}_p\{\widehat{\varepsilon}(k,\beta),s\} = \frac{\sum_{t=1}^T \widehat{\varepsilon}(k,\beta)_t^2}{\widehat{\sigma}(k,\beta)^2} - T + 2 \cdot s
\end{equation}
with $\widehat{\sigma}(k,\beta)^2$ the variance of $\widehat{\varepsilon}(k,\beta)$. $C_p\{\widehat{\varepsilon}(k,\beta),s\}$ tends to choose models which overfit and is not consistent in selecting the true model, see \textcite{mallick_bayesian_2013}, \textcite{woodroofe_model_1982} and \textcite{nishii_asymptotic_1984}.

The FPE uses the formula
	\begin{equation}
	\text{FPE}\{\widehat{\varepsilon}(k,\beta),s\} = \frac{T + s}{(T - s)T}\sum_{t=1}^T \widehat{\varepsilon}(k,\beta)_t^2,
	\end{equation}
see \textcite{akaike_statistical_1970}

So far, the discussed criteria depend on little data information. Just the squared residuals and, in the case of Mallows' C$_p$, the variance are taken into account. The AIC uses more information by depending on the maximum likelihood, derived by
\begin{equation}
L\{\widehat{\varepsilon}(k,\beta)\} = \max_\beta \prod_t f\{\widehat{\varepsilon}(k,\beta)_t\},
\end{equation}
where $f$, in (\ref{formula_Eepsilon}), represents the density of the $\widehat{\varepsilon}(k,\beta)_{t}$ over all $t$. The AIC is defined to be
\begin{equation}
\label{formula_AIC}
\text{AIC}\{\widehat{\varepsilon}(k,\beta),s\} = -2 \log L\{\widehat{\varepsilon}(k,\beta)\} + s \cdot 2,
\end{equation}
\textcite{akaike_information_1998}. If the true model is of finite dimension, then the AIC is not consistent, compare \textcite{hurvich_regression_1989}. \textcite{shibata_asymptotic_1983} showed the asymptotic efficiency of Mallows' C$_p$ and AIC under the assumption of an infinite number of regression variables or an increasing number of regression variables with the sample size. Due to the usage of the density in deriving the AIC, it uses more information about the dataset. Considering that (\ref{formula_min_eps}) implies the criteria are derived under an expected squared loss function, 
	\begin{equation}
	\label{formula_Eepsilon}
	{\changefont{cmss}{m}{n} \operatorname{\text{E}}} (\|\varepsilon(k,\beta)\|^2) = \int_{-\infty}^{\infty} \| \varepsilon(k,\beta) \|^2_2 f\{\varepsilon(k,\beta)\} d\varepsilon(k,\beta),
	\end{equation}
the density, $f$, can be estimated different from the Gaussian distribution. Here, $f$  
is estimated nonparametrically with an Epanechnikov kernel, since according to \textcite{hardle_nonparametric_2004} the \textcite{epanechnikov_non-parametric_1969} kernel shows a good balance between variance optimization and numerical performance. In nonparametric estimation with an Epanechnikov kernel, Epa, the estimator of $f$ is derived by
\begin{equation*}
\widehat{f}_h(x) = \frac{1}{nh} \sum_{i=1}^{n} \text{Epa}(\frac{x - x_i}{h}), \quad \text{Epa}(u) = \frac{3}{4\sqrt{5}} (1 - \frac{u^2}{5}) \Ind(|u| \leq \sqrt{5})
\end{equation*}
where $h$ is the bandwidth.

The bandwidth selection is performed with the plug-in selector by \textcite{sheather_reliable_1991} and further described in \textcite{wand_multivariate_1994}. The plug-in selector is derived under the loss function Mean Integrated Squared Error, MISE. \textcite{hall_kullback-leibler_1987} found that the Kullback-Leibler (KL) loss function for selecting the smoothing parameter of the kernel density is highly influenced by the tails of the distribution. \textcite{devroye_nonparametric_1985} mention that Mean Integrated Error (MIE) is stronger affected than MISE by the tails of the distribution and \textcite{kanazawa_hellinger_1993} claims that MIE shall be used if interest is in modeling the tails. \textcite{kanazawa_hellinger_1993} investigates that the use of a Kullback-Leibler loss function would put more weight on the tails compared to MISE. Since this is not in our interest, the choice of the density smoothing parameter, $h$, is performed under MISE. 

Due to the richer information basis of the AIC, we decide to use it as the selection criteria for CRIX. The choice is supported by an empirical analysis in section \ref{Chapter5}.

To decide with the AIC which number $k$ should be used, a procedure was created which compares the squared difference between log returns of the TMI, see Definition \ref{def_TotalMarketIndex}, and several candidate indices,
\begin{equation}
\label{epsilon_hat}
\| \widehat{\varepsilon}(k_j,\beta) \|^2 = \| \varepsilon(k_{max})^{TMI} - \varepsilon(k_j,\beta)^{CRIX} \|^2,
\end{equation}
where $\varepsilon(k_j,\beta)^{CRIX}$ is the log return of CRIX version with $k_j$ constituents and $\widehat{\varepsilon}(k_j,\beta)$ is the respective difference. 
The candidate indices, $\text{CRIX}(k_j,\beta)$, have different numbers of constituents which fulfill $k_1 < k_2 < k_3 < \cdots$, where $k_j = k_1 + s (j-1)$. Therefore, the number of constituents between the indices are equally spaced. The procedure implies that the selection method evaluates if $s$ more assets add information to CRIX. If so, these assets are added to the intercept and the next $s$ assets are tested for. Assets with a higher market capitalization are expected to have a higher influence on the AIC, so the following theorem is formulated: 
\begin{theorem}
	\label{theorem_size_asset}
	The rate of improvement of the AIC depends on the relative value of an asset in the market.
\end{theorem}
The proof for the Theorem \ref{theorem_size_asset} is given in the Appendix, \ref{Appendix_AIC}, under the assumption of normally distributed error terms. Therefore, we will follow the common practise to include the assets with the highest market capitalization in the index,
\begin{equation}
\underset{i}{\arg\max} \sum_{j=1}^{k} P_{j,i,t^-_l} Q_{j,i,t^-_l}, \quad i \in \{1, \dots, K\}.
\end{equation}
Thus, a top-down approach to decide about the number of index constituents is applied.

For the sorting of the index constituents by highest market capitalization, just the closing data of the last day of a month are used. We chose to do so, since the next periods CRIX will just depend on $Q_{i,t^-_l}$, (\ref{formula_CRIX}), and not on data which lie further in the past. This is in line with the methodology of e.g. the DAX. For LCRIX, the CCs with the highest trading volume are chosen respectively,
\begin{equation}
\underset{i}{\arg\max} \sum_{j=1}^{k} Vol_{j,i,t^-_l}, \quad i \in \{1, \dots, K\}.
\end{equation}

Since the differences between the $\text{TMI}(k_{max})$ and $\text{CRIX}(k_j,\beta)$ are caused over time by the missing time series in $\text{CRIX}(k_j,\beta)$, the independence assumption of the $\widehat{\varepsilon}(k_j,\beta)$ for all $j$ can not be fulfilled by construction. But \textcite{gyorfi_nonparametric_1989} give arguments that under certain conditions in case of nonparametric density estimation, the rate of convergence is essentially the same as for an independent sample. 
Summarizing the described procedure, results to:
\begin{enumerate}
	\item At time point $T + 1$, construct $\text{TMI}(k_{max})$
	\item Set $j = 2$
	\item Construct $\text{CRIX}(k_1,1) \text{ and } \text{CRIX}(k_j,\beta), \,\, k_1 < k_2 < k_3 < \cdots$
	\item Compute $\widehat{\varepsilon}(k_j, \beta)$ and $\widehat{\varepsilon}(k_1, 1)$
	\item Kernel density estimation (KDE) for density 
	$f(\widehat{\varepsilon}(k_1, 1))$ 
	\begin{enumerate}
		\item Compute the log likelihood (\ref{formula_AIC}) for $\widehat{\varepsilon}(k_j, \beta)$ with KDE for $\widehat{\varepsilon}(k_1, 1)$.
		\item Sum the log likelihoods
	\end{enumerate}
	\item Derive AIC$\{\widehat{\varepsilon}(k_j, \beta),k_j-k_1\}$ and AIC$\{\widehat{\varepsilon}(k_1, 1),0\}$
	\item If $j = (k_{max} - k_1)/k_1$: stop, else jump to 3. and $j = j + 1$
\end{enumerate}

The next section describes the further index rules for CRIX.

\section{CRIX family rules}
\label{Chapter3}
The constituents of the indices are regularly checked so that the corresponding index always represents its asset universe well. It is common to do this on a quarterly basis. In case of CRIX this reallocation is much faster. In the past, coins have shown a very volatile behavior, not just in the manner of price volatility. In some weeks, many occur out of nothing in the market and many others vanish from the market even when they were before very important, e.g., Auroracoin. This calls for a faster reallocation of the market benchmark than on a quarterly basis. A monthly reallocation is chosen to make sure that CRIX catches the momentum of the CC market well. Therefore, on the last day of every month, the CCs which had the highest market capitalization on the last day in the last month will be checked and the first $k$ will be included in CRIX for the coming month. Accordingly for LCRIX the ones with the highest trading volume are chosen. 

Since a review of an index is commonly performed on a quarterly basis the number of index members of CRIX will be checked on a quarterly basis too. The described procedure from Section \ref{Chapter2} will be applied to the observations from the last three months on the last day of the third month after the markets closed. The number of index constituents, $k$, will be used for the next three months. Thus, CRIX corresponds to a monthly rebalanced portfolio which number of constituents is reviewed quarterly.

It may happen that some data are missing for some of the analyzed time series. 
If an isolated missing value occurs alone in the dataset, meaning that the values before and after it are not missing, then Missing At Random (MAR) is assumed. This assumption means that just observed information cause the missingness, \textcite{horton_much_2007}. The Last-Observation-Carried-Forward (LOCF) method is then applied to fill the gap for the application of the AIC. We did not choose a different approach since a regression or imputation method may alter the data in the wrong direction. By LOCF, no change is implied and the CC is not excluded. If two or more data are missing in a row, then the MAR assumption may be violated, therefore no method is applied. The corresponding time series is then excluded from the computation in the derivation period. If data are missing during the computation of the index values, the LOCF method is applied too. This is done to make the index insensitive to this CC at this time point. CRIX should mimic market changes, therefore an imputation or regression method for the missing data would distort the view on the market.

Before continuing, the described rules are summarized:
\begin{itemize}
	\item Quarterly altering of the number of index constituents
	\item Monthly altering of the index constituents
	\item Model selection for index derivation with AIC
	\item Nonparametric estimation of the density
	\item Application of a top-down approach to select the assets for the subset analysis
	\item Application of LOCF if trading of an asset stops before next reallocation.
\end{itemize}

\section{The CRIX family}
\label{Chapter4}
Using the described methods and rules from above, three indices will be proposed. This indices provide a different look at the market. 

\begin{enumerate}
	\item CRIX/LCRIX: \\The first and leading index is CRIX and for volume weighting LCRIX. While the choice for the best number of constituents is made, their numbers are chosen in steps of 5. It is common in financial industry to construct market indices with a number of constituents which is evenly divisible by 5, see e.g. \textcite{ftse_ftse_2016}, \textcite{s&p_index_2014}, \textcite{deutsche_boerse_ag_guide_2013}. Therefore this selection is applied for CRIX$(k)$, $k = 5, 10, 15, \dots$ with $k_1 = 5$. Since the global minimum for the selection criterion may involve many index constituents, but a sparse index is the goal, the search for the optimal model terminates at level $j$ whenever 
	\begin{equation}
	\text{AIC}\{\widehat{\varepsilon}(k_j, \beta),k_j-5\} < \text{AIC}\{\widehat{\varepsilon}(k_{j-1}, \beta),k_{j-1}-5\}
	\end{equation}
	and $k_{j-1}$ index constituents are chosen.
	Therefore merely a local optimum will be achieved in most of the cases for $\Theta = \Theta_{AIC}$, in (\ref{equation_Theta_SC}). But the choice is still asymptotically optimal by defining $\Theta = \{\Theta_{AIC} | k_i \leq k_j \forall i\}$. In Section \ref{Chapter5} it will be shown that the performance of the index is already very good.
	\item ECRIX/LECRIX: \\The second constructed index is called Exact CRIX (ECRIX) and Liquidity ECRIX respectively. It follows the above rules too. 
	But the number of its constituents is chosen in steps of 1. Therefore the set of models contains CRIX$(k)$, $k = 1, 2, 3, \dots$ with $k_1 = 1$ and stops when
	\begin{equation}
	\text{AIC}\{\widehat{\varepsilon}(k_{j}, \beta),k_{j}-1\} < \text{AIC}\{\widehat{\varepsilon}(k_{j-1}, \beta),k_{j-1}-1\}.
	\end{equation}
	\item EFCRIX/LEFCRIX: \\Since the decision procedures for CRIX and ECRIX terminate when the AIC rises for the first time, Exact Full CRIX and Liquidity EFCRIX will be constructed to visualize whether the decision procedure works fine for the covered indices. The intention is to have an index which may approach the TMI but only in case even small assets help improve the view on the total market, a benchmark for the benchmarks. It'll be derived with the AIC procedure, compare Section \ref{Chapter2}. For $k = 1, 2, 3, \dots$ with $k_1 = 1$ the decision rule is based on 
	\begin{equation}
	\min_{k_j, \beta} \text{AIC}\{\widehat{\varepsilon}(k_j, \beta),k_j-1\}
	\end{equation}
	for $\Theta = \Theta_{AIC}$, in (\ref{equation_Theta_SC}).
	This index computes the AIC for every possible number of constituents and the number is chosen where the AIC becomes minimal. 
\end{enumerate}

\section{Performance analysis}
\label{Chapter5}
The indices CRIX, ECRIX, EFCRIX with market cap weighting and LCRIX, LECRIX, LEFCRIX with volume weighting have been proposed to give insight into the CC market. Our RDC CC database covers data for over 1000 CCs, kindly provided by CoinGecko. The data used for the analysis cover daily closing data for prices, market volume and market capitalization in USD for each CC in the time period from 2014-04-01 to 2017-03-25. Crypto exchanges are open on the weekends, therefore data for weekend closing prices exist. Since CC exchanges do not finish trading after a certain time point every day, a time point which serves as a closing time has to be defined. CoinGecko used 12 am UTC time zone. One should note that missing data are observed in the dataset, therefore the last rules from Chapter \ref{Chapter3} will come into play.

\begin{figure}
	\begin{center}
		\includegraphics[scale = 0.5]{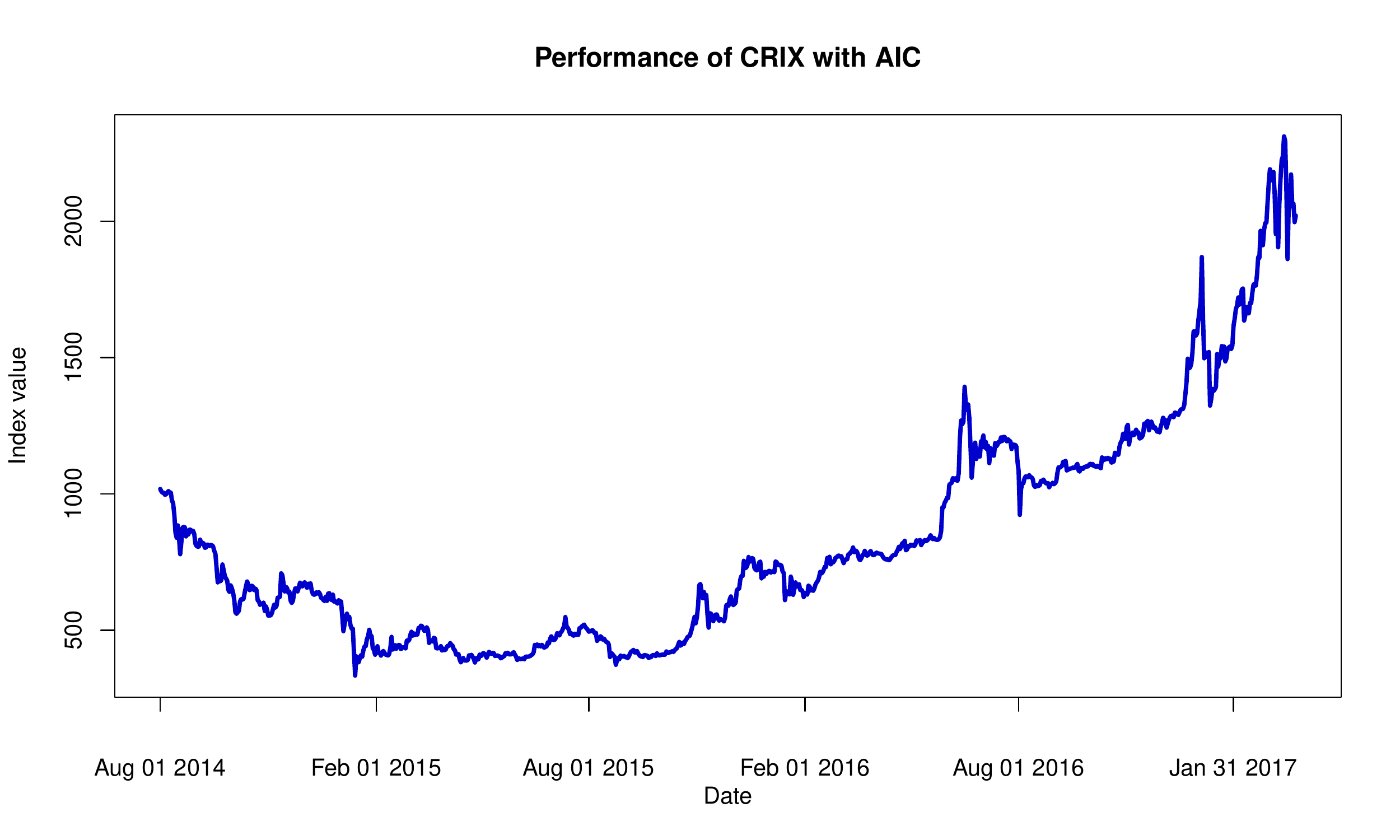}
		\caption{Performance of \textcolor{blue3}{CRIX} }\hfill\raisebox{-1pt}{\includegraphics[scale=0.05]{quantlet} \href{https://github.com/QuantLet/CRIX}{CRIXindex}} \raisebox{-1pt}{\includegraphics[scale=0.05]{quantlet}\href{https://github.com/QuantLet/CRIX}{CRIXcode}}
		\label{plot_crix}
	\end{center}
\end{figure}

\begin{figure}
	\begin{center}
		\includegraphics[scale = 0.5]{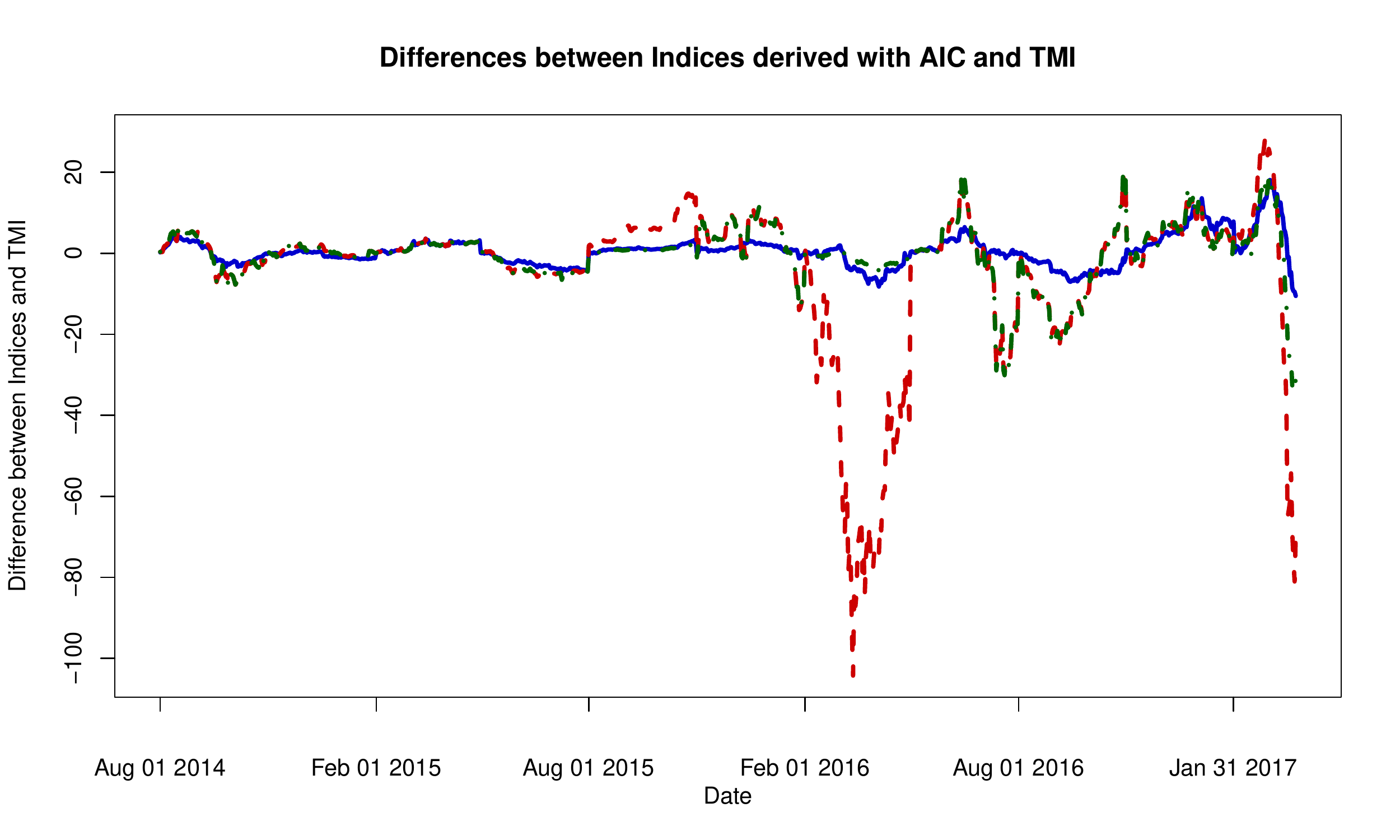}
		\caption{Realized difference between TMI and \textcolor{blue3}{CRIX} (solid), \textcolor{red3}{ECRIX} (dashed), \textcolor{darkgreen}{EFCRIX} (dotdashed)} \hfill\raisebox{-1pt}{\includegraphics[scale=0.05]{quantlet} \href{https://github.com/QuantLet/CRIX}{CRIXfamdiff}}
		\raisebox{-1pt}{\includegraphics[scale=0.05]{quantlet}\href{https://github.com/QuantLet/CRIX}{CRIXcode}}
		\label{plot_CRIXECRIXEFCRIXTotalMarket}
	\end{center}
\end{figure}

Figure \ref{plot_crix} shows the performance of CRIX, and Figure \ref{plot_CRIXECRIXEFCRIXTotalMarket} the differences between CRIX and both ECRIX and EFCRIX. For the purpose of comparison, the indices were recalibrated on the recalculation dates since the index constituents change then. We do not provide each index plot individually since they perform almost equally. However, the AIC method gave very different numbers of constituents for the corresponding indices. The numbers of constituents are given in Table \ref{table_IC_membersOverall}. For comparison, the number of constituents under the other discussed model selection criteria are provided too. The variance of $C_p$ was derived with a GARCH(1,1) model, \textcite{bollerslev_generalized_1986}. The corresponding information for ECRIX and EFCRIX are given in the same Table, \ref{table_IC_membersOverall}. Interestingly the methodology of EFCRIX causes its number of constituents to reach a relatively stable value for each period. ECRIX has mostly much fewer constituents than CRIX and EFCRIX due to the fact that this index just runs until a local optimum. Comparing the number of constituents for CRIX derived with AIC against the other criteria, one sees that GC, GFC and SH tend to choose more or the same number of constituents than AIC. Also all three criteria suggest the same result. $C_p$ stops at the initial value for CRIX, ECRIX and EFCRIX. For CRIX, ECRIX and EFCRIX, AIC mostly chooses less constituents compared to all other criteria, except $C_p$ which terminates very early. For LCRIX, LECRIX and LEFCRIX mostly less constituents were chosen than for CRIX, ECRIX and EFCRIX, compare Table \ref{table_IC_membersOverallL}. Note that the AIC gave the sparsest result again.

% latex table generated in R 3.3.0 by xtable 1.8-2 package
% Sun Aug 13 08:29:22 2017
\begin{table}[ht]
\centering
\begin{tabular}{l|rrrrrr}
  \hline
\hline
 & AIC & GC & GFC & SH & Cp & FPE \\ 
  \hline
CRIX & 0.4769 & 0.4883 & 0.3755 & 0.3598 & 1.9844 & 0.0042 \\ 
  ECRIX & 11.0988 & 10.3673 & 10.3673 & 10.4667 & 79.3979 & 0.0048 \\ 
  EFCRIX & 3.1394 & 0.0116 & 0.0049 & 0.0049 & 79.3979 & 0.0048 \\ 
  LCRIX & 0.6417 & 0.1497 & 0.1217 & 0.1211 & 0.6638 & 0.0049 \\ 
  LECRIX & 22.8782 & 16.7187 & 16.7187 & 16.7187 & 125.0620 & 0.0047 \\ 
  LEFCRIX & 7.9158 & 0.0645 & 0.0126 & 0.0126 & 125.0620 & 0.0047 \\ 
  btc & 79.3979 & 79.3979 & 79.3979 & 79.3979 & 79.3979 & 79.3979 \\ 
   \hline
\hline
\end{tabular}
\caption{Comparison of CRIX, ECRIX, EFCRIX, derived under different penalizations, against TMI under mean of monthly Mean Squared Error, compared with btc} 
\label{table_mse}
\end{table}

% latex table generated in R 3.3.0 by xtable 1.8-2 package
% Sun Aug 13 08:29:22 2017
\begin{table}[ht]
\centering
\begin{tabular}{l|rrrrrr}
  \hline
\hline
 & AIC & GC & GFC & SH & Cp & FPE \\ 
  \hline
CRIX & 0.9896 & 0.9908 & 0.9918 & 0.9928 & 0.9835 & 1.0000 \\ 
  ECRIX & 0.9576 & 0.9586 & 0.9586 & 0.9586 & 0.9133 & 1.0000 \\ 
  EFCRIX & 0.9794 & 0.9990 & 1.0000 & 1.0000 & 0.9133 & 1.0000 \\ 
  LCRIX & 0.9928 & 0.9949 & 0.9959 & 0.9959 & 0.9917 & 1.0000 \\ 
  LECRIX & 0.9692 & 0.9700 & 0.9700 & 0.9700 & 0.9501 & 1.0000 \\ 
  LEFCRIX & 0.9855 & 0.9979 & 1.0000 & 1.0000 & 0.9501 & 1.0000 \\ 
  btc & 0.9133 & 0.9133 & 0.9133 & 0.9133 & 0.9133 & 0.9133 \\ 
   \hline
\hline
\end{tabular}
\caption{Comparison of CRIX, ECRIX, EFCRIX, derived under different penalizations, against TMI under mean of monthly Mean Directional Accuracy, compared with btc} 
\label{table_mda}
\end{table}

The indices optimized until a local optimum are expected to perform less optimal than the globally optimized ones against the TMI/LTMI. Table \ref{table_mse} and Table \ref{table_mda} give the mean over monthly Mean Squared Error (MSE) and Mean Directional Accuracy (MDA), defined as
\begin{alignat}{2}
\text{MSE}\{\text{CRIX}(k)\} &= \rlap{$\displaystyle\frac{1}{t_{l}^{+} - t_{l}^{-}} \sum_{t = t_{l}^{-}}^{t_{l}^{+}} \{\text{CRIX}(k)_t - \text{TMI}(k_{\max})_t\}^2$} \\
\text{MDA}\{\text{CRIX}(k)\} &= \frac{1}{t_{l}^{+} - t_{l}^{-}} \sum_{t=t_{l}^{-}}^{t_{l}^{+}} \Ind[&&\mbox{sign}\{\text{TMI}(k_{\max})_t - \text{TMI}(k_{\max})_{t-1}\} \nonumber \\
& &&= \mbox{sign}\{\text{CRIX}(k)_t - \text{CRIX}(k)_{t-1}\}]
\end{alignat}
where $t_{l}^{-}$ and $t_{l}^{+}$ are the beginning and end of the month respectively, $\Ind(\cdot)$ is the indicator function and $\mbox{sign}(\cdot)$ gives the sign of the respective equation. 
Apparently CRIX performs best, which can be explained due to its larger number of index constituents. The CRIX, ECRIX and EFCRIX are close in terms of the MDA but the MSE is much better for CRIX. Comparing all the model selection criteria, FPE has the best performance in terms of MSE and MDA, due to choosing high numbers of constituents. The trading volume weighted indices are close in terms of MSE and MDA to their market weighted corresponding indices. At the same time the number of constituents are mostly sparser for the volume weighted ones. 

\begin{figure}
	\begin{center}
		\includegraphics[scale = 0.5]{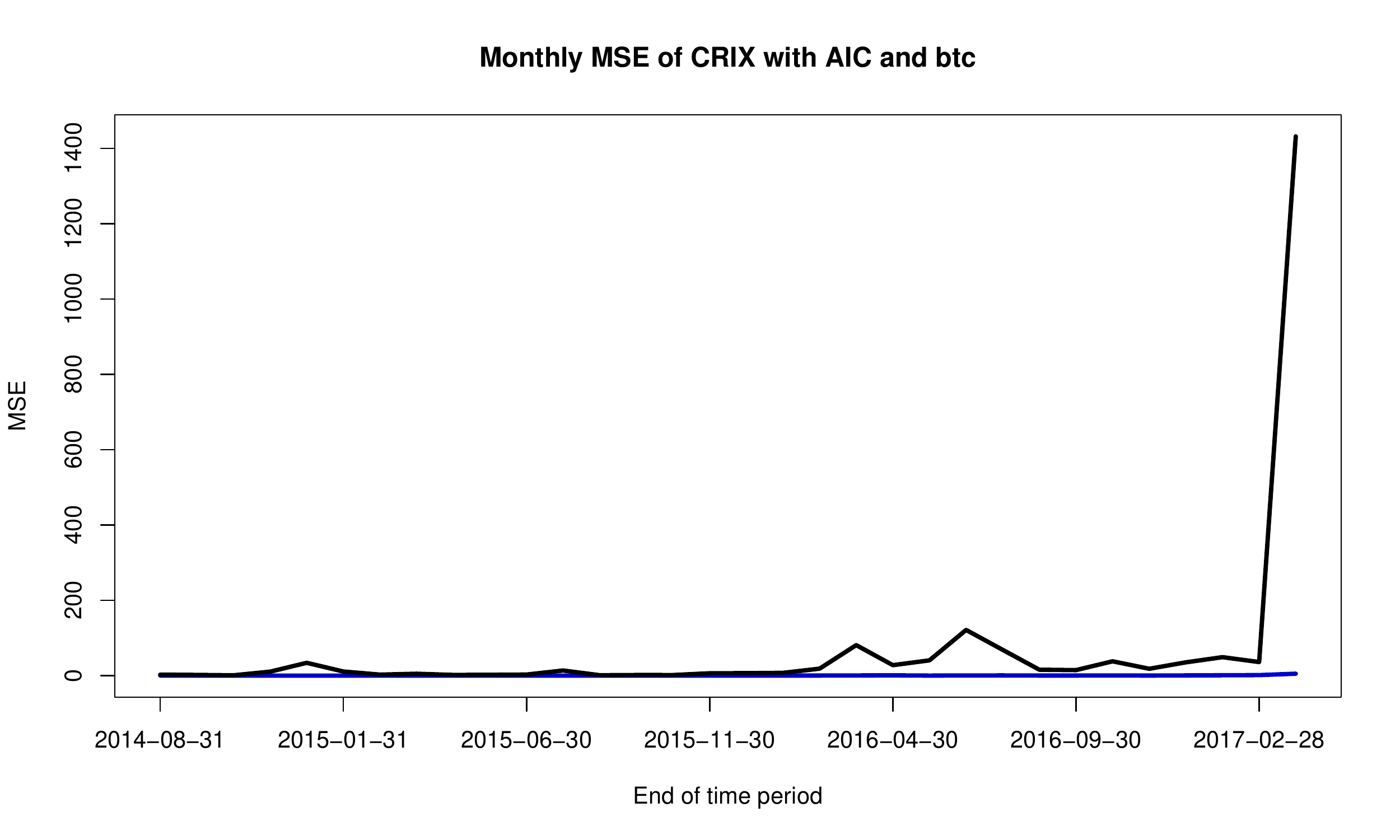}
		\caption{Performance of \textcolor{blue3}{CRIX} compared to BTC}
		\label{plot_CRIXagainstBTC}
	\end{center}
\end{figure}

CRIX was constructed with steps of five which is common in practice and performed best under AIC. For this case the number of constituents was the most stable, while achieving the best performance for MSE and MDA. Additionally, the analysis showed that it is indeed unnecessary from a practical viewpoint to choose the global optimal AIC under steps of 1. Even a local optimum and a much more stable number of constituents is able to mimic the market movements very well in terms of the MDA and MSE. Furthermore, even for ECRIX there was more than one constituent selected most of the time. This shows that Bitcoin, which currently clearly dominates the market in terms of market capitalization and trading volume, does not account for all the variance in the market. Other CCs are important for the market movements too.

Depending on the theoretical and empirical analysis, we decided to continue with the AIC. From the theoretical viewpoint, the AIC uses the most information about the data, since it relies on the density. From the empirical analysis, the AIC chooses much less constituents than GC, GFC, SH and FPE, while its performance in terms of MSE and MDA is close to the three outlined criteria. The better performance was achieved due to overparametrization of the index by GC, GFC, SH and FPE. Therefore, CRIX will be derived with the AIC criterion.

\begin{figure}
	\centering
	\begin{subfigure}{0.45\textwidth}
		\includegraphics[height = 0.2\textheight, width=1\textwidth]{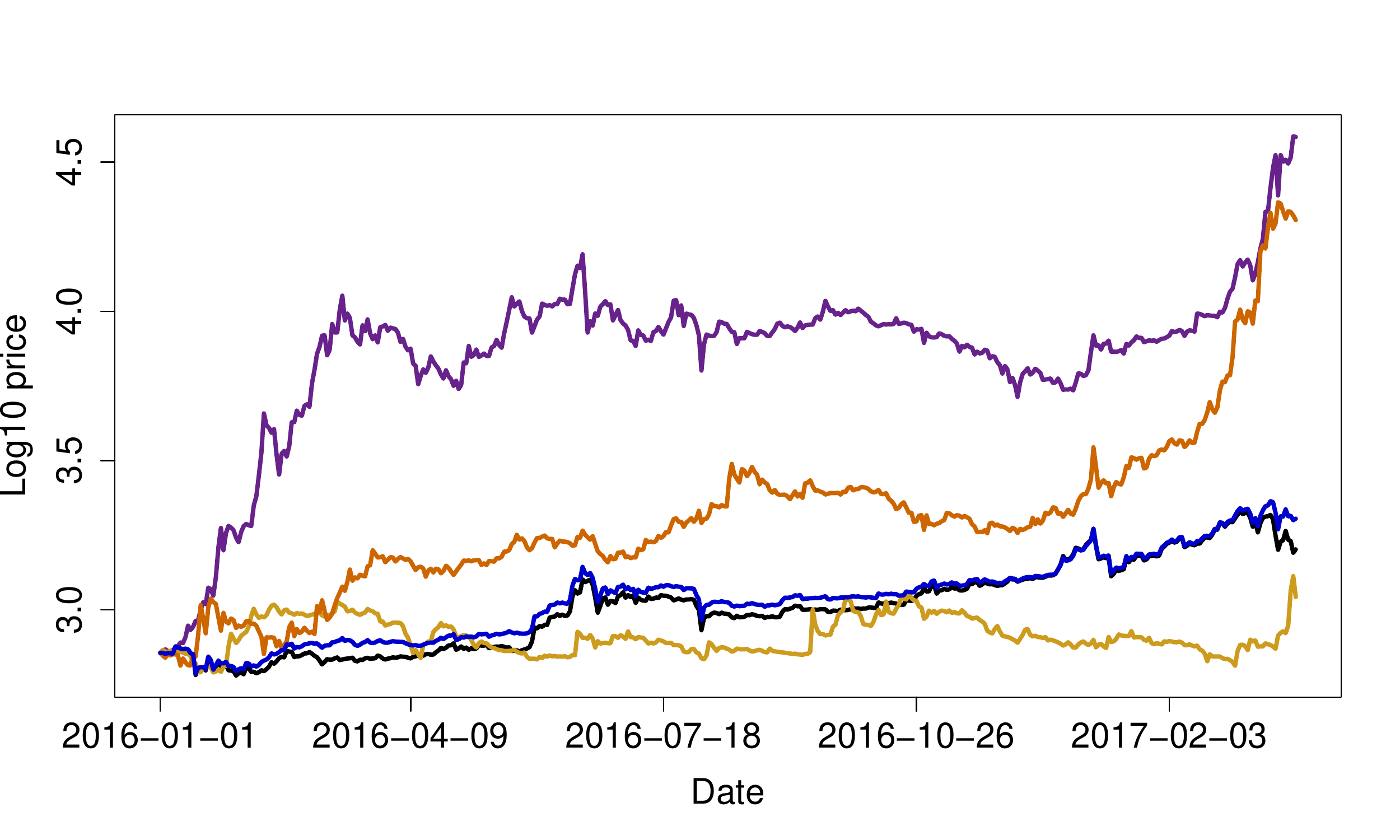}
		\caption{Performance of rescaled log price series of \textcolor{blue3}{CRIX}, BTC, \textcolor{darkorchid4}{ETH}, \textcolor{goldenrod3}{XRP} and \textcolor{darkorange3}{DASH}}\label{CompareCRIXBtcEthXrpDash}
	\end{subfigure}	
	\begin{subfigure}{0.45\textwidth}
		\includegraphics[height = 0.2\textheight, width=1\textwidth]{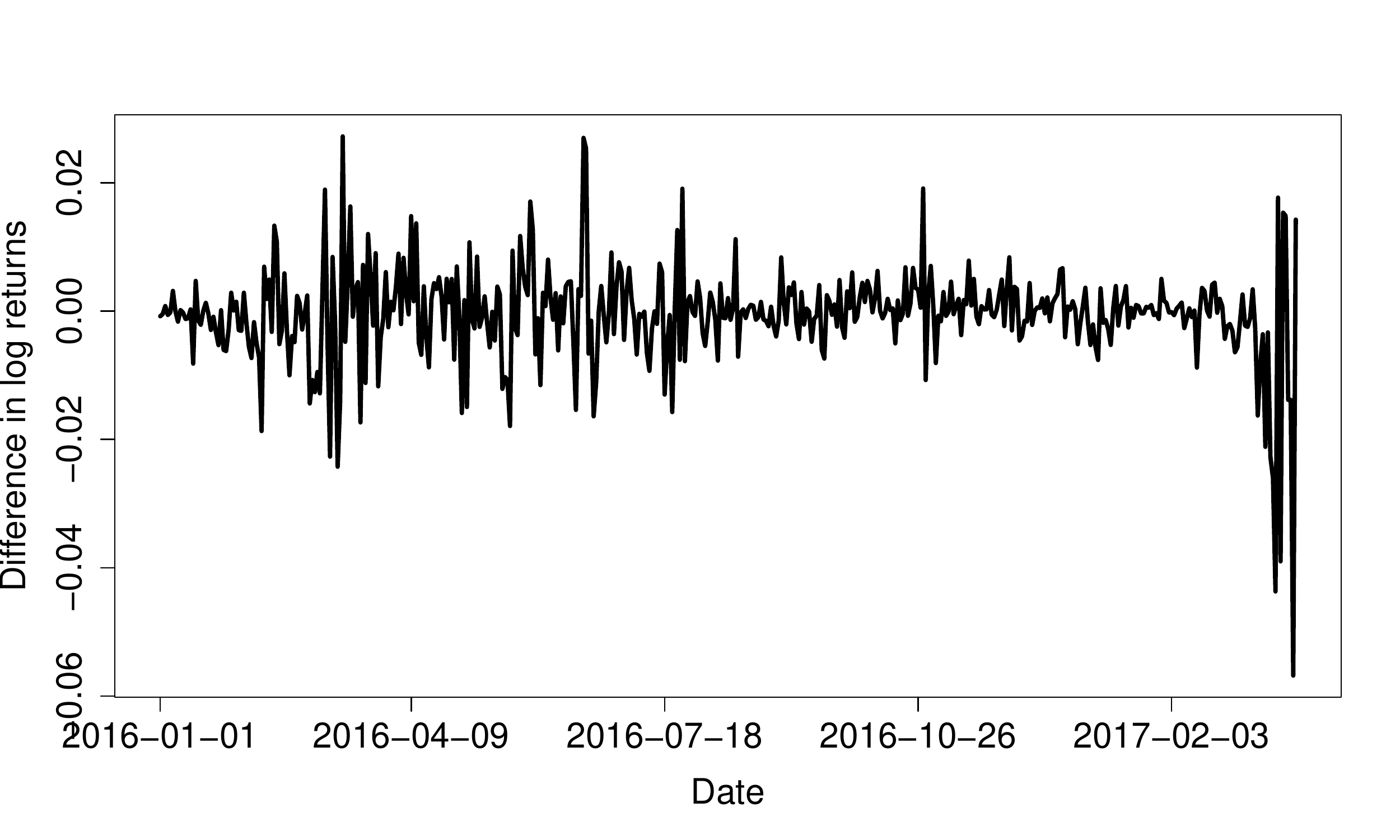}
		\caption{Difference in the log returns of \textcolor{blue3}{CRIX} and BTC}\label{CompareCRIXBtc}
	\end{subfigure}	
\caption{Comparison of performance of CRIX, BTC, ETH, XRP and DASH}\label{CompareDouble}
\end{figure}

\begin{figure}
	\centering
	\begin{subfigure}{0.45\textwidth}
		\includegraphics[height = 0.2\textheight, width=1\textwidth]{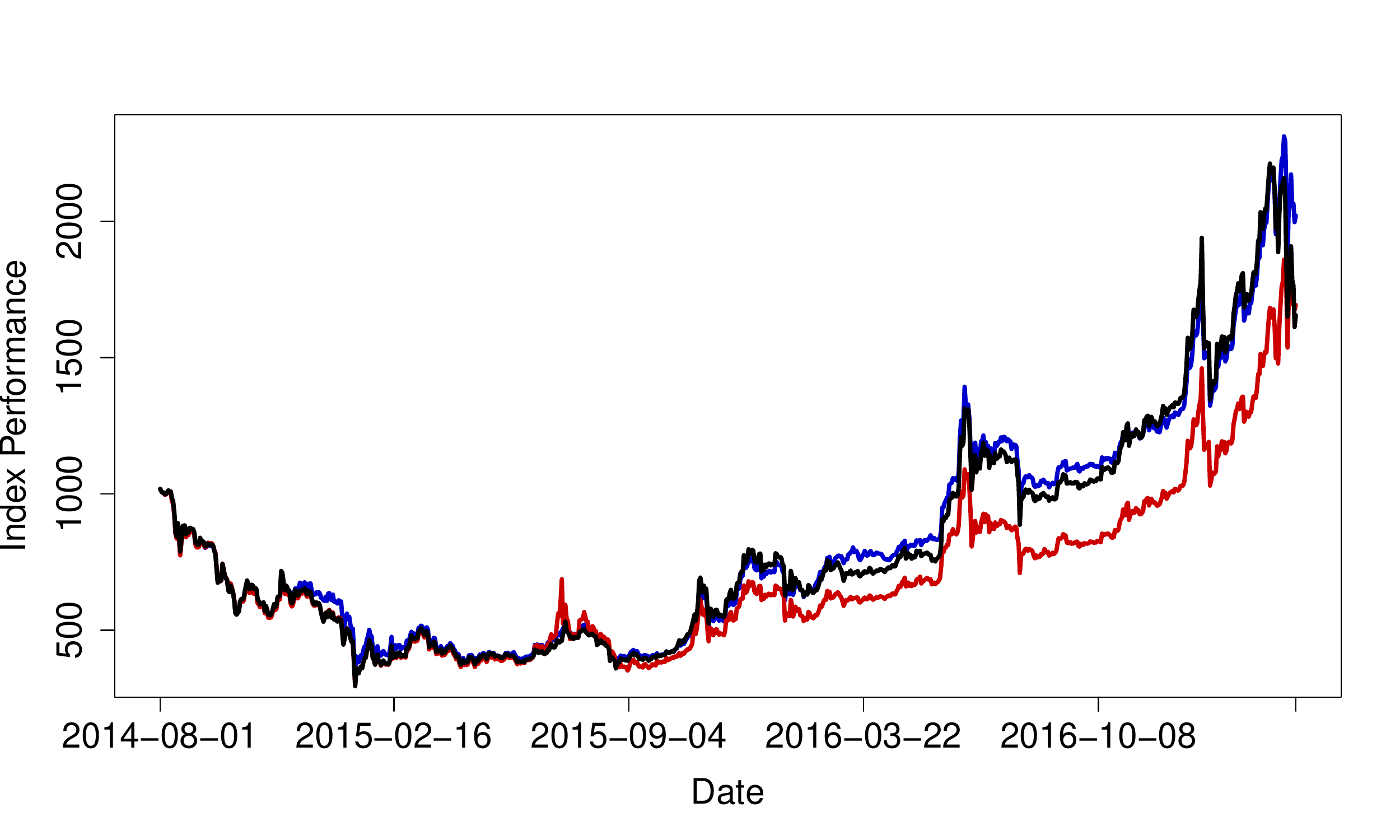}
		\caption{Performance of index series of \textcolor{blue3}{CRIX}, BTC and \textcolor{red3}{LCRIX}}\label{CompareCRIXLCRIXBtc}
	\end{subfigure}	
	\begin{subfigure}{0.45\textwidth}
		\includegraphics[height = 0.2\textheight, width=1\textwidth]{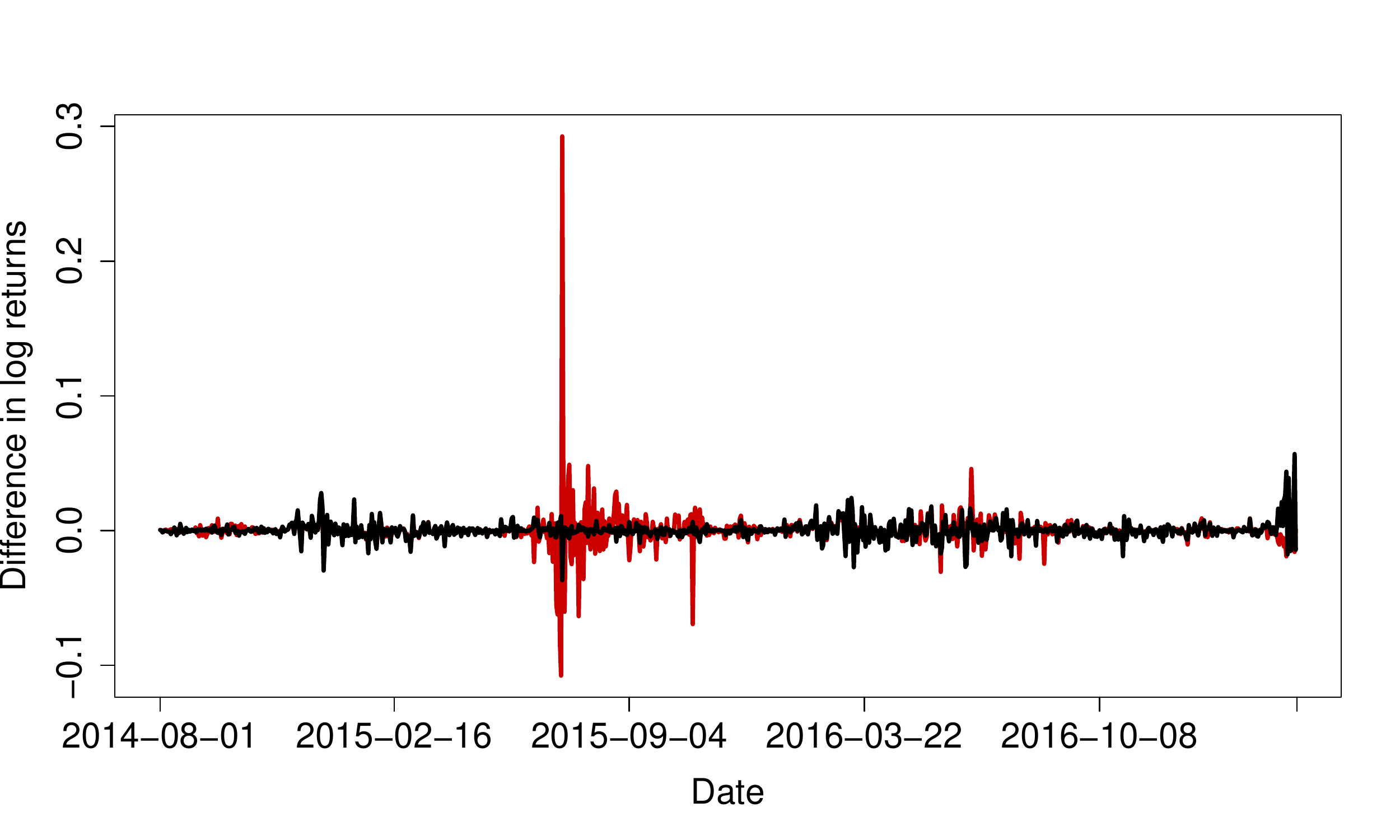}
		\caption{Difference in the log returns of \textcolor{red3}{CRIX and LCRIX} and CRIX and BTC}\label{CompareDiffLogCRIXLCRIX}
	\end{subfigure}	
	\caption{Comparison of performance of \textcolor{blue3}{CRIX}, BTC and \textcolor{red3}{LCRIX}}\label{CompareDoubleDiffLog}
\end{figure}

Comparing CRIX with the development of BTC, it tracks the market development better over time. Figure \ref{plot_CRIXagainstBTC} shows the monthly MSE of CRIX with AIC and BTC. In 2016 CRIX tracked the market development much better than BTC, and in the beginning of 2017 even better due to the huge impact of the price gain of altcoins like Ethereum, Ripple and Dash. Their performance is visualized in Figure \ref{CompareDouble}, clearly showing the better performance of CRIX in this time period, driven by price gains in altcoins. Due to the log scale and the high gains of altcoins, the difference between CRIX and BTC appears little, while in fact being considerable. Figure \ref{CompareCRIXBtc} shows the difference in the log returns of CRIX and BTC. One sees differences in their return series, which are particularly strong beginning of 2016 and in March 2017. Comparing the performance of CRIX and LCRIX against BTC, one observes an increasing spread between the indices, Figure \ref{CompareDoubleDiffLog}. It indicates a lower weight of BTC in LCRIX, thus tackling the issue of dominance of BTC in CRIX by liquidity weighting. Having a look at the actual differences in the log return series compared to CRIX, Figure \ref{CompareDiffLogCRIXLCRIX}, stronger spikes are observed, thus showing the difference in the performance from CRIX and LCRIX driven by the stronger weights on altcoins in LCRIX. Table \ref{table_weights} shows the actual weights given to BTC and altcoins in the respective indices. In the liquidity indices altcoins frequently receive a higher weight compared to the respective indices based on market capitalization weighting. Once the altcoins received even $52\%$ of the weights in LCRIX. The results show the market focus in terms of trading is stronger for altcoins than their market capitalization suggests, thus an index accounting for this is called for, LCRIX. Simultaneously the weighting scheme tackles the dominance of BTC in a market capitalization index.

\addtolength{\tabcolsep}{-3pt} 
% latex table generated in R 3.3.0 by xtable 1.8-2 package
% Sun Aug 13 08:29:22 2017
\begin{table}[ht]
\centering
\begingroup\footnotesize
\begin{tabular}{l|rrrrrrrrrrrrrrrrrrr}
  \hline
\hline
  & \multicolumn{6}{c}{CRIX}& \multicolumn{6}{c}{ECRIX}& \multicolumn{6}{c}{EFCRIX}&\\& AIC & GC & GFC & SH & Cp & FPE & AIC & GC & GFC & SH & Cp & FPE & AIC & GC & GFC & SH & Cp & FPE & max\\ \hline
1 & 5 & 10 & 10 & 10 & 10 & 35 & 2 & 2 & 2 & 3 & 1 & 36 & 2 & 7 & 30 & 30 & 1 & 36 & 36 \\ 
  2 & 10 & 15 & 15 & 15 & 5 & 100 & 3 & 3 & 3 & 3 & 1 & 93 & 3 & 94 & 93 & 93 & 1 & 93 & 113 \\ 
  3 & 5 & 10 & 35 & 35 & 5 & 100 & 5 & 5 & 5 & 5 & 1 & 93 & 5 & 94 & 93 & 93 & 1 & 93 & 158 \\ 
  4 & 10 & 10 & 10 & 40 & 5 & 95 & 3 & 3 & 3 & 3 & 1 & 90 & 3 & 91 & 90 & 90 & 1 & 90 & 182 \\ 
  5 & 10 & 20 & 20 & 20 & 5 & 100 & 2 & 4 & 4 & 4 & 1 & 93 & 12 & 94 & 93 & 93 & 1 & 93 & 169 \\ 
  6 & 10 & 10 & 20 & 20 & 5 & 100 & 2 & 2 & 2 & 2 & 1 & 93 & 2 & 94 & 93 & 93 & 1 & 93 & 171 \\ 
  7 & 5 & 20 & 20 & 20 & 5 & 100 & 1 & 1 & 1 & 1 & 1 & 93 & 16 & 94 & 93 & 93 & 1 & 93 & 176 \\ 
  8 & 15 & 20 & 20 & 20 & 5 & 95 & 3 & 4 & 4 & 4 & 1 & 91 & 3 & 92 & 91 & 91 & 1 & 91 & 140 \\ 
  9 & 15 & 5 & 5 & 5 & 5 & 100 & 3 & 3 & 3 & 3 & 1 & 93 & 3 & 94 & 93 & 93 & 1 & 93 & 188 \\ 
  10 & 15 & 15 & 25 & 25 & 5 & 100 & 3 & 5 & 5 & 5 & 1 & 93 & 3 & 94 & 93 & 93 & 1 & 93 & 207 \\ 
  11 & 10 & 35 & 45 & 45 & 5 & 100 & 2 & 2 & 2 & 2 & 1 & 93 & 4 & 94 & 93 & 93 & 1 & 93 & 221 \\ 
   \hline
\hline
\end{tabular}
\endgroup
\caption{Comparison of AIC, GC, GFC, SH, Cp and the FPE method for the selection of the number of index constituents for the CRIX, ECRIX and EFCRIX in the 11 periods} 
\label{table_IC_membersOverall}
\end{table}

\addtolength{\tabcolsep}{3pt} 

\textcolor{red}{
\addtolength{\tabcolsep}{-3pt} 
% latex table generated in R 3.3.0 by xtable 1.8-2 package
% Sun Aug 13 08:29:22 2017
\begin{table}[ht]
\centering
\begingroup\footnotesize
\begin{tabular}{l|rrrrrrrrrrrrrrrrrrr}
  \hline
\hline
  & \multicolumn{6}{c}{LCRIX}& \multicolumn{6}{c}{LECRIX}& \multicolumn{6}{c}{LEFCRIX}&\\& AIC & GC & GFC & SH & Cp & FPE & AIC & GC & GFC & SH & Cp & FPE & AIC & GC & GFC & SH & Cp & FPE & max\\ \hline
1 & 5 & 10 & 15 & 15 & 5 & 35 & 2 & 3 & 3 & 3 & 1 & 36 & 2 & 6 & 16 & 16 & 1 & 36 & 36 \\ 
  2 & 5 & 10 & 10 & 10 & 5 & 100 & 2 & 4 & 4 & 4 & 1 & 93 & 2 & 94 & 93 & 93 & 1 & 93 & 113 \\ 
  3 & 5 & 5 & 20 & 20 & 5 & 100 & 3 & 4 & 4 & 4 & 1 & 93 & 6 & 94 & 93 & 93 & 1 & 93 & 158 \\ 
  4 & 15 & 20 & 20 & 30 & 5 & 95 & 3 & 2 & 2 & 2 & 1 & 90 & 3 & 91 & 90 & 90 & 1 & 90 & 182 \\ 
  5 & 5 & 5 & 5 & 5 & 5 & 100 & 1 & 1 & 1 & 1 & 1 & 93 & 93 & 94 & 93 & 93 & 1 & 93 & 169 \\ 
  6 & 5 & 5 & 5 & 5 & 5 & 100 & 2 & 2 & 2 & 2 & 1 & 93 & 9 & 94 & 93 & 93 & 1 & 93 & 171 \\ 
  7 & 10 & 25 & 30 & 35 & 5 & 100 & 1 & 2 & 2 & 2 & 1 & 93 & 1 & 94 & 93 & 93 & 1 & 93 & 176 \\ 
  8 & 10 & 20 & 35 & 35 & 5 & 95 & 1 & 1 & 1 & 1 & 1 & 91 & 3 & 92 & 91 & 91 & 1 & 91 & 140 \\ 
  9 & 5 & 10 & 10 & 10 & 5 & 100 & 2 & 2 & 2 & 2 & 1 & 93 & 2 & 94 & 93 & 93 & 1 & 93 & 188 \\ 
  10 & 10 & 10 & 10 & 10 & 5 & 100 & 1 & 1 & 1 & 1 & 1 & 93 & 1 & 94 & 93 & 93 & 1 & 93 & 207 \\ 
  11 & 5 & 15 & 15 & 15 & 5 & 100 & 2 & 3 & 3 & 3 & 1 & 93 & 2 & 94 & 93 & 93 & 1 & 93 & 221 \\ 
   \hline
\hline
\end{tabular}
\endgroup
\caption{Comparison of AIC, GC, GFC, SH, Cp and the FPE method for the selection of the number of index constituents for the LCRIX, LECRIX and LEFCRIX in the 11 periods} 
\label{table_IC_membersOverallL}
\end{table}

\addtolength{\tabcolsep}{3pt} 
}

\section{Application to the German stock market}
\label{Chapter6}
The CRIX methodology was derived with the idea of finding a method which allows mimicking young and fast changing markets appropriately. But well known major markets usually change their structure too. So the proposed methodology is tested on the German stock market, which has four major indices: DAX, MDAX, SDAX and TecDAX. The DAX is used to determine the overall market direction, \textcite{jansen_deutsche_1992}. Since it is chosen from the so called prime segment, it has some prior restrictions. It is interesting to see whether our methodology yields the DAX as an adequate benchmark for the total market. Since the indices are derived with market cap weighting scheme, only this methodology is tested. Following Definition \ref{def_TotalMarketIndex}, all available stocks are defined as the TMI and our new method is applied to find an appropriate index. Again, the 7-step method from Section \ref{Chapter2} was applied to find the number of constituents, but it starts at 30 members to check if more constituents are necessary. The method for the identification of $k$ and the reallocation of the included assets is performed quarterly, like DAX. 
To be in line with the DAX reallocation dates, the index calculation will start after the third Friday of September and the reallocation dates are the third Fridays of December, March, June and September, see \textcite{deutsche_boerse_ag_guide_2013}.

The data were fetched from Datastream in the period 2000-06-16 until 2015-12-18. All stocks which are German companies and are traded on XETRA are chosen. Any time series for which Datastream reported an error either for the price or market capitalization data was excluded from the analysis. The index, computed with the new methodology, is called Flexible DAX (FDAX). One should note that the analysis starts three months after the starting point of the dataset due to the initialization period of FDAX.

\begin{figure}
	\begin{center}
		\includegraphics[scale = 0.5]{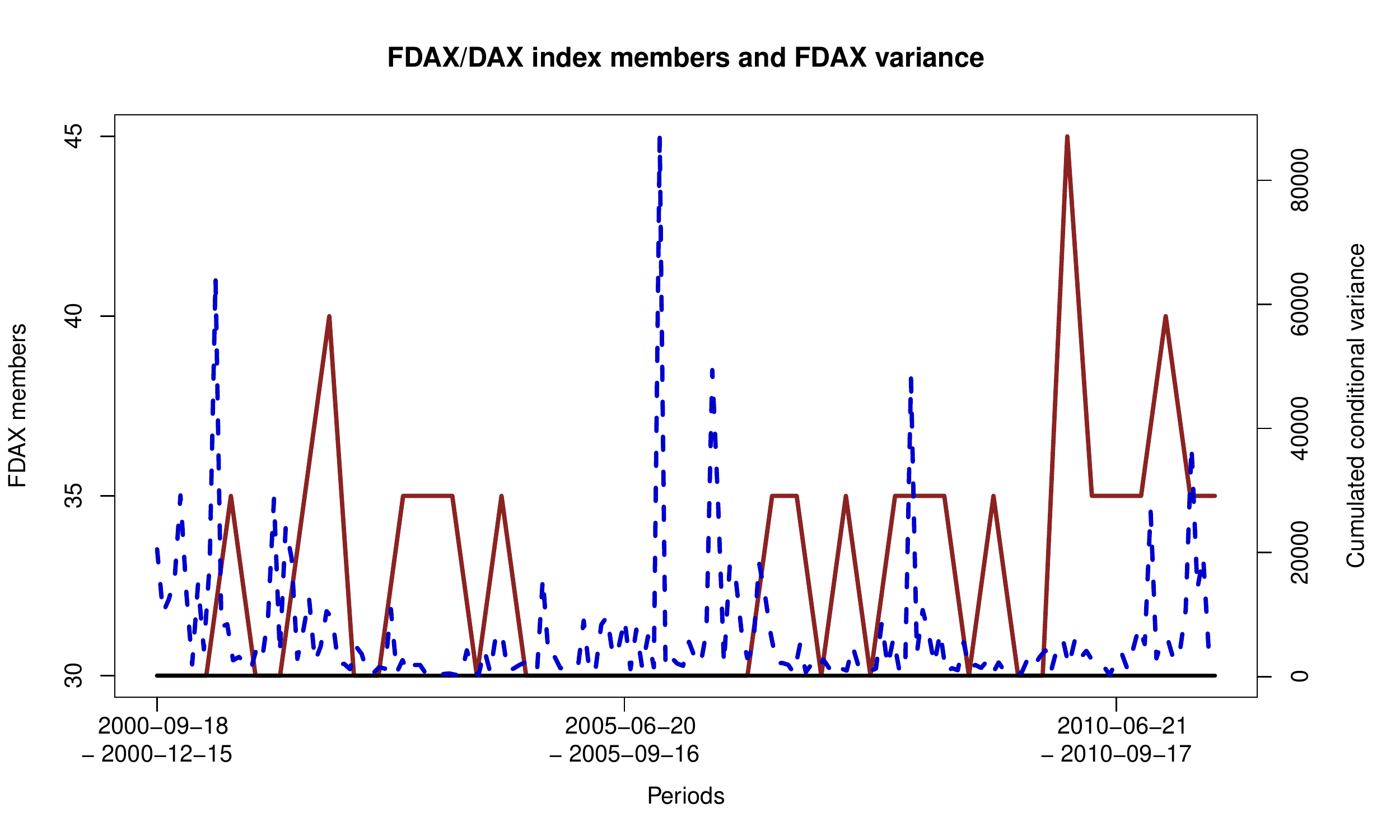}
		\caption{Number of constituents of \textcolor{brown4}{FDAX} (solid), DAX (horizontal solid) and \textcolor{blue3}{cumulated monthly variance of FDAX} (dashed)} \hfill\raisebox{-1pt}{\includegraphics[scale=0.05]{quantlet} \href{https://github.com/QuantLet/CRIX}{CRIXdaxmembersvar}}
		\raisebox{-1pt}{\includegraphics[scale=0.05]{quantlet}\href{https://github.com/QuantLet/CRIX}{CRIXcode}}
		\label{plot_DAXCRIXComparisonMembersGARCH}
	\end{center}
\end{figure}

Figure \ref{plot_DAXCRIXComparisonMembersGARCH} shows the number of members of FDAX and DAX in the respective periods. Most of the time, the number of index constituents for FDAX is higher than the 30 members of DAX. Just around 2004-2005 is the $k$ more frequently 30. Especially while the turmoil of the financial markets, starting from 2008/2009, is the number of index constituents much higher.
One might hint that a higher reported variability in one period should cause an increase in $k$ in the next period, since it was shown that the selection method depends on the variance, see Section \ref{Chapter9}. Figure \ref{plot_DAXCRIXComparisonMembersGARCH} shows that this idea can partially be supported. The derivation of the conditional variance was performed with a GARCH(1,1) model, \textcite{bollerslev_generalized_1986}, and the daily results were summed up. Obviously, in the extreme cases increases the $k$ in the next period, see 2001, 2002, 2006 and 2011.

The computation of the MSE and MDA, see Table \ref{table_mse_mda_daxcrix}, shows that FDAX is a more accurate benchmark for the total market as DAX. Since \textcite{jansen_deutsche_1992} state that DAX may be used to analyze the movements of the total market, an MDA of 92 \% is indeed good.  
But FDAX mimics the market even better, with an MDA of 96 \%. Also the MSE for FDAX is much lower than the one of DAX. Therefore the methodology fulfilled its goal to find a sparse, investable and accurate benchmark, depending on the MDA. 

% latex table generated in R 3.3.0 by xtable 1.8-2 package
% Sun Aug 13 08:28:20 2017
\begin{table}[ht]
\centering
\begin{tabular}{l|rr}
  \hline
\hline
 & MSE & MDA \\ 
  \hline
FDAX vs. TMI & 6.36 & 0.96 \\ 
  DAX vs. TMI & 51.02 & 0.92 \\ 
   \hline
\hline
\end{tabular}
\caption{Comparison of DAX with CRIX methodology (FDAX) and rescaled DAX against TMI} 
\label{table_mse_mda_daxcrix}
\end{table}

\section{Application to Mexican stock market}
\label{Chapter6.1}
The Mexican stock market is represented by the IPC35, \textcite{mexbol_prices_2013}. One of its rules is a readjustment of the weights to lower the effect of dominant stocks. In the CC market BTC is such a dominant asset. The CRIX methodology could help to circumvent arbitrary rules and develop an index to represent the market accurately. 

\begin{figure}
	\begin{center}
		\includegraphics[scale = 0.5]{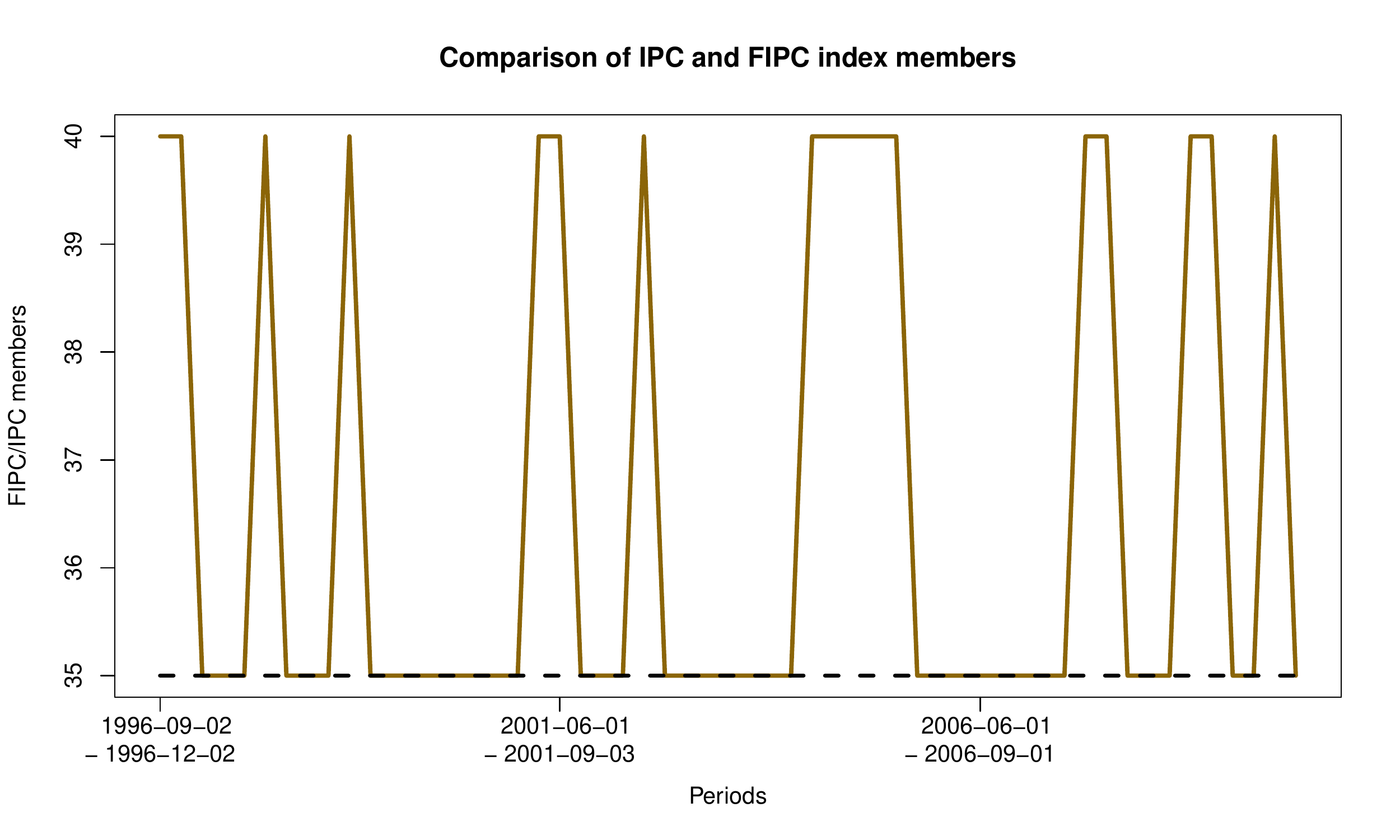}
		\caption{Number of constituents of \textcolor{darkgoldenrod4}{FIPC} (solid) and IPC (dashed) in the respective periods} \hfill\raisebox{-1pt}{\includegraphics[scale=0.05]{quantlet} \href{https://github.com/QuantLet/CRIX}{CRIXipcmembers}}
		\raisebox{-1pt}{\includegraphics[scale=0.05]{quantlet}\href{https://github.com/QuantLet/CRIX}{CRIXcode}}
		\label{plot_IPCCRIXIPCComparisonMembers}
	\end{center}
\end{figure}

The data were fetched from Datastream for the period 1996-06-01 until 2015-05-29 and cover all Mexican companies listed in Datastream. The specifications of the methodology are the same as for the German stock market except for the recalculation date. In line with the methodology of the IPC35, the index is recalculated with the closing data of the last business days of August, November, February and May, therefore the recalculated index starts on the first business days of September, December, March and June. The TMI will be all fetched companies. The choice of $k$ starts with 35 since this is the amount of constituents of IPC.

Again, the CRIX methodology works well. The MSE is very low compared to the one for the IPC35 and the MDA gives a much better performance too, see Table \ref{table_mse_mda_ipccrix}. We can conclude that the methodology helped to circumvent the usage of arbitrary rules for the weights in the rules of the indices and enhances at the same time the performance of the market index. Figure \ref{plot_IPCCRIXIPCComparisonMembers} shows the number of index members of the FIPC compared to the IPC. Obviously, the methodology also suggests using more than 35 index members half of the time which is the number of members of the IPC.

% latex table generated in R 3.3.0 by xtable 1.8-2 package
% Sun Aug 13 08:27:39 2017
\begin{table}[ht]
\centering
\begin{tabular}{l|rr}
  \hline
\hline
 & MSE & MDA \\ 
  \hline
FIPC vs. TMI & 24.97 & 0.97 \\ 
  IPC vs. TMI & 4743.50 & 0.91 \\ 
   \hline
\hline
\end{tabular}
\caption{Comparison of IPC with CRIX methodology (FIPC) and rescaled IPC against TMI} 
\label{table_mse_mda_ipccrix}
\end{table}

\section{Conclusion}
\label{Chapter8}
The movements of CCs are very different from each other, \textcite{elendner_cross-section_2017}. So studying the entire market of CCs requires an instrument which adequately captures and displays the market movements, an index. But index construction for CCs requires a new methodology to find the right number of index members. Innovative markets, like the one for CCs, change their structure frequently. The proposed methods were applied to oracle a new family of indices, which are displayed and updated on a daily basis. The performance of the new indices were studied and it was shown that the dynamic AIC based methodology results in indices with stable properties. 
The results show that a market like the CC market - momentarily dominated by Bitcoin - still needs a representative index since Bitcoin does not account for all the variance in the market. The diversified nature of the CC market makes the inclusion of altcoins in the index product critical to improve tracking performance. We have shown that assigning optimal weights to altcoins helps to reduce the tracking errors of a CC portfolio, despite the fact that their market cap is much smaller relative to Bitcoin. 

Besides the classical market capitalization weighting, a volume weighting scheme was proposed. The corresponding indices are sparser in terms of constituents while having a comparable performance, which gives support to this weighting scheme under the goals of the study. 
The AIC based method was also applied to the German stock market. The results yield a more accurate benchmark in terms of MDA. In applying the CRIX methodology to the Mexican stock market, which is dominated by Telmex, one finds high accuracy of it in terms of MSE and MDA.

We conclude, that the CRIX technology enhances the construction of an index if the goal is to find a sparse, investable and accurate benchmark.

\section{Acknowledgments}
We would like to thank the editor and an anonymous referee for their valuable comments to this article. Our thanks extends to David Lee Kuo Chuen and Ernie G. S. Teo for their comments in several discussions. Financial support from the Deutsche Forschungsgemeinschaft via CRC 649 ”Economic Risk” and
IRTG 1792 ”High Dimensional Non Stationary Time Series”, Humboldt-Universität zu Berlin, is gratefully acknowledged.

\printbibliography

@incollection{akaike_information_1998,
	series = {Springer {Series} in {Statistics}},
	title = {Information {Theory} and an {Extension} of the {Maximum} {Likelihood} {Principle}},
	copyright = {©1998 Springer-Verlag New York, Inc.},
	isbn = {978-1-4612-7248-9 978-1-4612-1694-0},
	urldate = {2015-11-16},
	booktitle = {Selected {Papers} of {Hirotugu} {Akaike}},
	publisher = {Springer New York},
	author = {Akaike, Hirotogu},
	editor = {Parzen, Emanuel and Tanabe, Kunio and Kitagawa, Genshiro},
	year = {1998},
	pages = {199--213}
}

@article{hall_kullback-leibler_1987,
	title = {On {Kullback}-{Leibler} {Loss} and {Density} {Estimation}},
	volume = {15},
	issn = {0090-5364},
	number = {4},
	urldate = {2016-08-31},
	journal = {The Annals of Statistics},
	author = {Hall, Peter},
	year = {1987},
	pages = {1491--1519}
}

@article{mexbol_prices_2013,
	title = {Prices and {Quotations} {Index} ({MEXBOL}) - {Methodology} {Note}},
	journal = {bmv.com},
	author = {{MEXBOL}},
	year = {2013}
}

@article{jansen_deutsche_1992,
	title = {Der {Deutsche} {Aktienindex} {DAX}},
	urldate = {2015-12-09},
	journal = {Fritz Knapp Verlag},
	author = {Janßen, Birgit and Rudolph, Bernd},
	year = {1992}
}

@incollection{ron_quantitative_2013,
	series = {Lecture {Notes} in {Computer} {Science}},
	title = {Quantitative {Analysis} of the {Full} {Bitcoin} {Transaction} {Graph}},
	copyright = {©2013 Springer-Verlag Berlin Heidelberg},
	isbn = {978-3-642-39883-4 978-3-642-39884-1},
	number = {7859},
	urldate = {2016-04-13},
	booktitle = {Financial {Cryptography} and {Data} {Security}},
	publisher = {Springer Berlin Heidelberg},
	author = {Ron, Dorit and Shamir, Adi},
	editor = {Sadeghi, Ahmad-Reza},
	month = apr,
	year = {2013},
	note = {DOI: 10.1007/978-3-642-39884-1\_2},
	pages = {6--24}
}

@article{kanazawa_hellinger_1993,
	title = {Hellinger distance and {Kullback}—{Leibler} loss for the kernel density estimator},
	volume = {18},
	issn = {0167-7152},
	doi = {10.1016/0167-7152(93)90022-B},
	number = {4},
	urldate = {2016-09-01},
	journal = {Statistics \& Probability Letters},
	author = {Kanazawa, Yuichiro},
	month = nov,
	year = {1993},
	pages = {315--321}
}

@article{woodroofe_model_1982,
	title = {On {Model} {Selection} and the {ARC} {Sine} {Laws}},
	volume = {10},
	issn = {0090-5364},
	number = {4},
	urldate = {2016-01-04},
	journal = {The Annals of Statistics},
	author = {Woodroofe, Michael},
	year = {1982},
	pages = {1182--1194}
}

@article{wang_buzz_2017,
	title = {Buzz {Factor} or {Innovation} {Potential}: {What} {Explains} {Cryptocurrencies}’ {Returns}?},
	volume = {12},
	issn = {1932-6203},
	shorttitle = {Buzz {Factor} or {Innovation} {Potential}},
	url = {http://journals.plos.org/plosone/article?id=10.1371/journal.pone.0169556},
	doi = {10.1371/journal.pone.0169556},
	number = {1},
	urldate = {2017-04-26},
	journal = {PLOS ONE},
	author = {Wang, Sha and Vergne, Jean-Philippe},
	month = jan,
	year = {2017},
	pages = {e0169556}
}

@article{mallick_bayesian_2013,
	title = {Bayesian {Methods} for {High} {Dimensional} {Linear} {Models}},
	volume = {1},
	issn = {2155-6180},
	doi = {10.4172/2155-6180.S1-005},
	journal = {Journal of Biometrics \& Biostatistics},
	author = {Mallick, Himel and Yi, Nengjun},
	month = jun,
	year = {2013},
	pmid = {24511433},
	pmcid = {PMC3914549}
}

@article{ftse_ftse_2016,
	title = {{FTSE} {UK} {Index} {Series}},
	journal = {www.ftse.com},
	author = {{FTSE}},
	month = may,
	year = {2016}
}

@article{crsp_crsp_2015,
	title = {{CRSP} {U}.{S}. {Equity} {Indexes} {Methodology} {Guide}},
	journal = {crsp.com/},
	author = {{CRSP}},
	month = jan,
	year = {2015}
}

@book{hayek_denationalization_1990,
	edition = {3. Edition},
	title = {Denationalization of {Money}: {An} {Analysis} of the {Theory} and {Practice} of {Concurrent} {Currencies}},
	publisher = {London: Institute of Economic Affairs},
	author = {Hayek, Friedrich A.},
	year = {1990}
}

@incollection{chen_econometric_2017,
	title = {Econometric {Analysis} of a {Cryptocurrency} {Index} for {Portfolio} {Investment}.},
	volume = {1},
	booktitle = {Handbook of {Digital} {Finance} and {Financial} {Inclusion}: {Cryptocurrency}, {FinTech}, {InsurTech}, and {Regulation}},
	publisher = {Elsevier},
	author = {Chen, Shi and Chen, Cathy Yi-Hsuan and Härdle, Wolfgang Karl and Lee, Teik Ming and Ong, Bobby},
	editor = {Lee Kuo Chuen, David and Deng, Robert},
	year = {2017}
}

@article{shibata_optimal_1981,
	title = {An {Optimal} {Selection} of {Regression} {Variables}},
	volume = {68},
	issn = {0006-3444},
	doi = {10.2307/2335804},
	number = {1},
	urldate = {2016-04-04},
	journal = {Biometrika},
	author = {Shibata, Ritei},
	year = {1981},
	pages = {45--54}
}

@article{mallows_comments_1973,
	title = {Some {Comments} on {Cp}},
	volume = {15},
	issn = {0040-1706},
	doi = {10.2307/1267380},
	number = {4},
	urldate = {2016-01-04},
	journal = {Technometrics},
	author = {Mallows, C. L.},
	year = {1973},
	pages = {661--675}
}

@incollection{reid_analysis_2013,
	title = {An {Analysis} of {Anonymity} in the {Bitcoin} {System}},
	copyright = {©2013 Springer Science+Business Media New York},
	isbn = {978-1-4614-4138-0 978-1-4614-4139-7},
	urldate = {2016-04-13},
	booktitle = {Security and {Privacy} in {Social} {Networks}},
	publisher = {Springer New York},
	author = {Reid, Fergal and Harrigan, Martin},
	editor = {Altshuler, Yaniv and Elovici, Yuval and Cremers, Armin B. and Aharony, Nadav and Pentland, Alex},
	year = {2013},
	note = {DOI: 10.1007/978-1-4614-4139-7\_10},
	pages = {197--223}
}

@article{epanechnikov_non-parametric_1969,
	title = {Non-{Parametric} {Estimation} of a {Multivariate} {Probability} {Density}},
	volume = {14},
	issn = {0040-585X},
	doi = {10.1137/1114019},
	number = {1},
	urldate = {2016-02-03},
	journal = {Theory of Probability \& Its Applications},
	author = {Epanechnikov, V.},
	month = jan,
	year = {1969},
	pages = {153--158}
}

@article{arlot_survey_2010,
	title = {A survey of cross-validation procedures for model selection},
	volume = {4},
	issn = {1935-7516},
	doi = {10.1214/09-SS054},
	urldate = {2016-04-14},
	journal = {Statistics Surveys},
	author = {Arlot, Sylvain and Celisse, Alain},
	year = {2010},
	mrnumber = {MR2602303},
	zmnumber = {1190.62080},
	pages = {40--79}
}

@article{hurvich_regression_1989,
	title = {Regression and time series model selection in small samples},
	volume = {76},
	issn = {0006-3444, 1464-3510},
	doi = {10.1093/biomet/76.2.297},
	number = {2},
	urldate = {2016-01-05},
	journal = {Biometrika},
	author = {Hurvich, Clifford M. and Tsai, Chih-Ling},
	month = jun,
	year = {1989},
	pages = {297--307}
}

@techreport{bolt_value_2016,
	address = {Rochester, NY},
	type = {{SSRN} {Scholarly} {Paper}},
	title = {On the {Value} of {Virtual} {Currencies}},
	url = {https://papers.ssrn.com/abstract=2842557},
	number = {ID 2842557},
	urldate = {2017-04-05},
	institution = {Social Science Research Network},
	author = {Bolt, Wilko and Oordt, Maarten van},
	month = sep,
	year = {2016}
}

@article{s&p_index_2014,
	title = {Index {Mathematics} - {Methodology}},
	journal = {us.spindices.com},
	author = {{S\&P}},
	year = {2014}
}

@incollection{elendner_cross-section_2017,
	title = {The {Cross}-{Section} of {Crypto}-{Currencies} as {Financial} {Assets}},
	volume = {1},
	booktitle = {Handbook of {Digital} {Finance} and {Financial} {Inclusion}: {Cryptocurrency}, {FinTech}, {InsurTech}, and {Regulation}},
	publisher = {Elsevier},
	author = {Elendner, Hermann and Trimborn, Simon and Ong, Bobby and Lee, Teik Ming},
	editor = {Lee Kuo Chuen, David and Deng, Robert},
	year = {2017}
}

@incollection{droge_comments_1996,
	series = {Contributions to {Statistics}},
	title = {Some {Comments} on {Cross}-{Validation}},
	copyright = {©1996 Physica-Verlag Heidelberg},
	isbn = {978-3-7908-0930-5 978-3-642-48425-4},
	urldate = {2016-04-05},
	booktitle = {Statistical {Theory} and {Computational} {Aspects} of {Smoothing}},
	publisher = {Physica-Verlag HD},
	author = {Droge, Bernd},
	editor = {Härdle, Wolfgang Karl and Schimek, Michael G.},
	year = {1996},
	note = {DOI: 10.1007/978-3-642-48425-4\_14},
	pages = {178--199}
}

@book{hardle_nonparametric_2004,
	title = {Nonparametric and {Semiparametric} {Models}},
	isbn = {978-3-642-17146-8},
	publisher = {Springer Science \& Business Media},
	author = {Härdle, Wolfgang Karl and Müller, Marlene and Sperlich, Stefan and Werwatz, Axel},
	year = {2004}
}

@article{kristoufek_what_2015,
	title = {What are the main drivers of the {Bitcoin} price? {Evidence} from wavelet coherence analysis},
	volume = {10},
	shorttitle = {What are the main drivers of the {Bitcoin} price?},
	urldate = {2015-08-21},
	journal = {PLOS ONE},
	author = {Kristoufek, Ladislav},
	month = apr,
	year = {2015},
	pages = {1--15}
}

@article{craven_smoothing_1978,
	title = {Smoothing noisy data with spline functions},
	volume = {31},
	issn = {0029-599X, 0945-3245},
	doi = {10.1007/BF01404567},
	number = {4},
	urldate = {2016-04-05},
	journal = {Numerische Mathematik},
	author = {Craven, Peter and Wahba, Grace},
	month = dec,
	year = {1978},
	pages = {377--403}
}

@article{shibata_asymptotic_1983,
	title = {Asymptotic mean efficiency of a selection of regression variables},
	volume = {35},
	issn = {0020-3157, 1572-9052},
	doi = {10.1007/BF02480998},
	number = {1},
	urldate = {2016-03-17},
	journal = {Annals of the Institute of Statistical Mathematics},
	author = {Shibata, Ritei},
	month = dec,
	year = {1983},
	pages = {415--423}
}

@book{devroye_nonparametric_1985,
	title = {Nonparametric {Density} {Estimation} {The} {L}1 {View}},
	publisher = {Wiley},
	author = {Devroye, Luc and Györfi, László},
	year = {1985}
}

@article{horton_much_2007,
	title = {Much ado about nothing: {A} comparison of missing data methods and software to fit incomplete data regression models},
	volume = {61},
	issn = {0003-1305},
	shorttitle = {Much ado about nothing},
	doi = {10.1198/000313007X172556},
	number = {1},
	journal = {The American Statistician},
	author = {Horton, Nicholas J. and Kleinman, Ken P.},
	month = feb,
	year = {2007},
	pmid = {17401454},
	pmcid = {PMC1839993},
	pages = {79--90}
}

@article{deutsche_boerse_ag_guide_2013,
	title = {Guide to the {Equity} {Indizes} of {Deutsche} {Boerse}},
	journal = {www.dax-indices.com},
	author = {{Deutsche Boerse AG}},
	year = {2013}
}

@techreport{white_market_2014,
	address = {Rochester, NY},
	type = {{SSRN} {Scholarly} {Paper}},
	title = {The {Market} for {Cryptocurrencies}},
	number = {ID 2538290},
	urldate = {2017-04-30},
	institution = {Social Science Research Network},
	author = {White, Lawrence H.},
	month = dec,
	year = {2014}
}

@article{s&p_dow_2015,
	title = {Dow {Jones} {Total} {Stock} {Market} {Indices} {Methodology}},
	journal = {us.spindices.com},
	author = {{S\&P}},
	month = jun,
	year = {2015}
}

@article{bollerslev_generalized_1986,
	title = {Generalized autoregressive conditional heteroskedasticity},
	volume = {31},
	issn = {0304-4076},
	doi = {10.1016/0304-4076(86)90063-1},
	number = {3},
	urldate = {2016-04-13},
	journal = {Journal of Econometrics},
	author = {Bollerslev, Tim},
	month = apr,
	year = {1986},
	pages = {307--327}
}

@article{econotimes_japans_2016,
	title = {Japans {Cabinet} {Approves} {New} {Bitcoin} {Regulations}},
	urldate = {2016-04-13},
	journal = {econotimes.com},
	author = {{EconoTimes}},
	month = mar,
	year = {2016}
}

@article{hardle_crix_2015,
	title = {{CRIX} or evaluating {Blockchain} based currencies},
	doi = {10.4171/OWR/2015/42},
	journal = {Oberwolfach Report No. 42/2015 ``The Mathematics and Statistics of Quantitative Risk''},
	author = {Härdle, Wolfgang Karl and Trimborn, Simon},
	year = {2015}
}

@article{nishii_asymptotic_1984,
	title = {Asymptotic {Properties} of {Criteria} for {Selection} of {Variables} in {Multiple} {Regression}},
	volume = {12},
	issn = {0090-5364, 2168-8966},
	doi = {10.1214/aos/1176346522},
	number = {2},
	urldate = {2015-11-17},
	journal = {The Annals of Statistics},
	author = {Nishii, Ryuei},
	month = jun,
	year = {1984},
	mrnumber = {MR740928},
	zmnumber = {0544.62063},
	pages = {758--765}
}

@article{kawa_bitcoin_2015,
	title = {Bitcoin {Is} {Officially} a {Commodity}, {According} to {U}.{S}. {Regulator}},
	urldate = {2016-04-13},
	journal = {Bloomberg.com},
	author = {Kawa, Luke},
	month = sep,
	year = {2015}
}

@book{gyorfi_nonparametric_1989,
	series = {Lecture {Notes} in {Statistics}},
	title = {Nonparametric {Curve} {Estimation} from {Time} {Series}},
	copyright = {©1989 Springer-Verlag Berlin Heidelberg},
	isbn = {978-0-387-97174-2 978-1-4612-3686-3},
	number = {60},
	urldate = {2015-12-14},
	publisher = {Springer New York},
	author = {Györfi, Lázió and Härdle, Wolfgang Karl and Sarda, Pascal and Vieu, Philippe},
	editor = {Györfi, Lázió and Härdle, Wolfgang Karl and Sarda, Pascal and Vieu, Philippe},
	year = {1989},
	note = {DOI: 10.1007/978-1-4612-3686-3}
}

@article{wand_multivariate_1994,
	title = {Multivariate plug-in bandwidth selection},
	volume = {9},
	issn = {0943-4062},
	number = {2},
	urldate = {2016-02-03},
	journal = {Computational Statistics},
	author = {Wand, M. P. and Jones, Michael Chris},
	year = {1994},
	pages = {97--116}
}

@article{sheather_reliable_1991,
	title = {A {Reliable} {Data}-{Based} {Bandwidth} {Selection} {Method} for {Kernel} {Density} {Estimation}},
	volume = {53},
	issn = {0035-9246},
	doi = {10.2307/2345597},
	journal = {Journal of the Royal Statistical Society. Series B. Methodological},
	author = {Sheather, Simon J. and Jones, Michael Chris},
	year = {1991},
	pages = {683--690}
}

@article{wilshire_associates_wilshire_2015,
	title = {Wilshire 5000 {Total} {Market} {Index} {Methodology}},
	journal = {wilshire.com},
	author = {{Wilshire Associates}},
	year = {2015}
}

@article{akaike_statistical_1970,
	title = {Statistical predictor identification},
	volume = {22},
	issn = {0020-3157, 1572-9052},
	doi = {10.1007/BF02506337},
	number = {1},
	urldate = {2016-01-05},
	journal = {Annals of the Institute of Statistical Mathematics},
	author = {Akaike, Hirotugu},
	month = dec,
	year = {1970},
	pages = {203--217}
}

\newpage
\section{Appendix}
\label{Chapter9}
\subsection{Proof of Theorem \ref{theorem_size_asset}}
\label{Appendix_AIC}

\textit{Proof:}
Assume normally distributed error terms, (\ref{formula_min_eps}) and (\ref{epsilon_hat}): $\varepsilon(k,\beta) \sim \mbox{N}\{0, \sigma(k,\beta)^2\}$, $\widehat{\varepsilon}(k,\beta) \sim \mbox{N}\{0, \hat{\sigma}(k,\beta)^2\}$. Then

\begin{equation}
\log L\{\varepsilon(k,\beta)\} = - \frac{T}{2} \log (2 \pi) - \frac{T}{2} \log \sigma(k,\beta)^2 - \frac{1}{2 \sigma(k, \beta)^2} \sum_{t = 1}^{T} \varepsilon(k,\beta)_t^2.
\end{equation}

Denote $RSS\{\widehat{\varepsilon}(k,\beta)\} = \sum_{t = 1}^{T} \widehat{\varepsilon}(k,\beta)_t^2$ and $\hat{\sigma}(k,\beta)^2 = T^{-1}RSS\{\widehat{\varepsilon}(k,\beta)\}$. Then

\begin{align}
\log L\{\widehat{\varepsilon}(k,\beta)\} &= - \frac{T}{2} \log (2 \pi) - \frac{T}{2} \log T^{-1}RSS\{\widehat{\varepsilon}(k,\beta)\} - \frac{1}{2 T^{-1}RSS\{\widehat{\varepsilon}(k,\beta)\}} RSS\{\widehat{\varepsilon}(k,\beta)\} \\
&= - \frac{T}{2} \log (2 \pi) - \frac{T}{2} \log T^{-1}RSS\{\widehat{\varepsilon}(k,\beta)\} - \frac{T}{2} \\
&= - \frac{T}{2} \log T^{-1}RSS\{\widehat{\varepsilon}(k,\beta)\} + C
\end{align}

with $C = - \frac{T}{2} \log (2 \pi) - \frac{T}{2}$. Since $C$ does not depend on any model parameters, just on the data length $T$, this part of the equation could be omitted. 

\begin{align}
\text{AIC}\{\widehat{\varepsilon}(k,\beta),s\} &= T \log T^{-1}RSS\{\widehat{\varepsilon}(k,\beta)\} + 2 \cdot s \\
&= T \log \widehat{\sigma}(k,\beta)^2 + 2 \cdot s
\end{align}

The enhancement in the fit to the Total Market Index (TMI) by adding more constituents, $s$, determines the degree of improvement of the likelihood. 

With the linearity property of the expectation operator, assume without loss of generality 
\begin{align*}
\E\{\varepsilon(k_{max})^{TM}\} = \E\{\varepsilon(k,\beta)^{CRIX}\} &= 0 \\
t &\in \{1, \dots, T\} \\
t_l^{-} &= 0 \\
s &= 1
\end{align*}

\begin{alignat}{2}
\widehat{\sigma}(k,\beta) &= \rlap{$\displaystyle\text{Var}\{\widehat{\varepsilon}(k, \beta)\}$} \nonumber\\
&= \rlap{$\displaystyle\text{Var}\{\varepsilon(k_{max})^{TM} - \varepsilon(k,\beta)^{CRIX}\}$} \nonumber\\
&= \sum_{t=1}^{T} \biggl[&& \log \left\{ \sum_{i=1}^{k_{max}} P_{it} Q_{i,0} ( \sum_{i=1}^k P_{i,t-1} Q_{i,0} + \beta_1 P_{k+1,t-1} Q_{k+1,0} ) \right\} \nonumber\\
& &&- \log \left\{ \sum_{i=1}^{k_{max}} P_{i,t-1} Q_{i,0} ( \sum_{i=1}^k P_{i,t} Q_{i,0} + \beta_1 P_{k+1,t} Q_{k+1,0} ) \right\} \biggr]^2\nonumber\\
&= \sum_{t=1}^{T} \biggl[&& \log \left\{ \sum_{i=1}^{k_{max}} P_{it} Q_{i,0} \sum_{i=1}^k P_{i,t-1} Q_{i,0} + \sum_{i=1}^{k_{max}} P_{it} Q_{i,0} \beta_1 P_{k+1,t-1} Q_{k+1,0} \right\} \nonumber\\
& &&- \log \left\{ \sum_{i=1}^{k_{max}} P_{i,t-1} Q_{i,0} \sum_{i=1}^k P_{i,t} Q_{i,0} + \sum_{i=1}^{k_{max}} P_{i,t-1} Q_{i,0} \beta_1 P_{k+1,t} Q_{k+1,0} \right\} \biggr]^2 \nonumber\\
\rlap{$\displaystyle\text{Using the relation } \log(a + b) = \log(a) + \log(1 + \frac{b}{a}) \text{, it results:}$} \nonumber\\
&= \sum_{t=1}^{T} \biggl[&& \log \left\{ \sum_{i=1}^{k_{max}} P_{it} Q_{i,0} \sum_{i=1}^k P_{i,t-1} Q_{i,0} \right\} + \log\left\{1 + \frac{ \sum_{i=1}^{k_{max}} P_{it} Q_{i,0} \beta_1 P_{k+1,t-1} Q_{k+1,0} }{ \sum_{i=1}^{k_{max}} P_{it} Q_{i,0} \sum_{i=1}^k P_{i,t-1} Q_{i,0} } \right\} \nonumber\\
& &&- \log \left\{ \sum_{i=1}^{k_{max}} P_{i,t-1} Q_{i,0} \sum_{i=1}^k P_{i,t} Q_{i,0} \right\} + \log \left\{1 + \frac{ \sum_{i=1}^{k_{max}} P_{i,t-1} Q_{i,0} \beta_1 P_{k+1,t} Q_{k+1,0} }{ \sum_{i=1}^{k_{max}} P_{i,t-1} Q_{i,0} \sum_{i=1}^k P_{i,t} Q_{i,0} } \right\} \biggr]^2 \nonumber\\
&= \sum_{t=1}^{T} \biggl(&& \log \left\{ \sum_{i=1}^{k_{max}} P_{it} Q_{i,0} \sum_{i=1}^k P_{i,t-1} Q_{i,0} \right\} - \log \left\{ \sum_{i=1}^{k_{max}} P_{i,t-1} Q_{i,0} \sum_{i=1}^k P_{i,t} Q_{i,0} \right\} \nonumber\\
& &&+ \left[ \log\left\{1 + \frac{ \beta_1 P_{k+1,t-1} Q_{k+1,0} }{ \sum_{i=1}^k P_{i,t-1} Q_{i,0} } \right\} - \log \left\{1 + \frac{ \beta_1 P_{k+1,t} Q_{k+1,0} }{ \sum_{i=1}^k P_{i,t} Q_{i,0} } \right\} \right] \biggr)^2 \label{proof_last_equ}
\end{alignat}
Solving the derivation and writing the terms which do not depend on $\beta_1$ as $A_t$ and the last part of (\ref{proof_last_equ}) as $B_t$:
\begin{align*}
\widehat{\sigma}(k,\beta) &= \sum_{t=1}^{T} A_t + 2 \log \left\{ \sum_{i=1}^{k_{max}} P_{it} Q_{i,0} \sum_{i=1}^k P_{i,t-1} Q_{i,0} \right\} B_t - 2 \log \left\{ \sum_{i=1}^{k_{max}} P_{i,t-1} Q_{i,0} \sum_{i=1}^k P_{i,t} Q_{i,0} \right\} B_t + B_t^2 \\
&= \sum_{t=1}^{T} A_t + 2 B_t \left[ \log \left\{ \sum_{i=1}^{k_{max}} P_{it} Q_{i,0} \sum_{i=1}^k P_{i,t-1} Q_{i,0} \right\} - \log \left\{ \sum_{i=1}^{k_{max}} P_{i,t-1} Q_{i,0} \sum_{i=1}^k P_{i,t} Q_{i,0} \right\} \right] + B_t^2 \\
&= \sum_{t=1}^{T} A_t + 2 B_t \left[ \varepsilon(k_{max})^{TM} - \varepsilon(k,1)^{CRIX} \right] + B_t^2 
\end{align*}
Since normally distributed error terms are assumed, note that $\beta_1 = \frac{Cov\{\widehat{\varepsilon}(k,1), \varepsilon_{k+1}\}}{Var\{\varepsilon_{k+1}\}}$, where $\varepsilon_{k+1}$ is the log return of $P_{i,t} Q_{i,0}$. The change in the variance will depend on the additional variance which the new constituent can explain, see $\beta_1$. Furthermore, it depends on the value of $P_{k+1,t} Q_{k+1,0}$ relative to $\sum_{i=1}^k P_{i,t} Q_{i,0}$, (\ref{proof_last_equ}), which is the summed market value of the constituents in the index. This infers that constituents with a higher market capitalization are more likely to be part of the index. 
$\blacksquare$

This gives support to using the often applied top-down approach, which we use for the construction of CRIX too.

% latex table generated in R 3.4.0 by xtable 1.8-2 package
% Wed Mar 14 18:47:39 2018
\begin{sidewaystable}[ht]
\centering
\begingroup\footnotesize
\begin{tabular}{l|rrrrrrrrrrrrrrrrrr}
  \hline
\hline
  Periods& \multicolumn{3}{c}{CRIX}& \multicolumn{3}{c}{LCRIX}& \multicolumn{3}{c}{ECRIX}& \multicolumn{3}{c}{LECRIX}& \multicolumn{3}{c}{EFCRIX}& \multicolumn{3}{c}{LEFCRIX}\\& k & BTC & altcoins & k & BTC & altcoins & k & BTC & altcoins & k & BTC & altcoins & k & BTC & altcoins & k & BTC & altcoins \\ \hline
2014/08 & 5 & 0.96 & 0.04 & 5 & 0.96 & 0.04 & 2 & 0.98 & 0.02 & 2 & 0.97 & 0.03 & 2 & 0.98 & 0.02 & 2 & 0.97 & 0.03 \\ 
  2014/09 & 5 & 0.94 & 0.06 & 5 & 0.82 & 0.18 & 2 & 0.97 & 0.03 & 2 & 0.86 & 0.14 & 2 & 0.97 & 0.03 & 2 & 0.86 & 0.14 \\ 
  2014/10 & 5 & 0.93 & 0.07 & 5 & 0.86 & 0.14 & 2 & 0.97 & 0.03 & 2 & 0.91 & 0.09 & 2 & 0.97 & 0.03 & 2 & 0.91 & 0.09 \\ 
  2014/11 & 10 & 0.92 & 0.08 & 5 & 0.95 & 0.05 & 3 & 0.94 & 0.06 & 2 & 0.99 & 0.01 & 3 & 0.94 & 0.06 & 2 & 0.99 & 0.01 \\ 
  2014/12 & 10 & 0.85 & 0.15 & 5 & 0.95 & 0.05 & 3 & 0.86 & 0.14 & 2 & 0.97 & 0.03 & 3 & 0.86 & 0.14 & 2 & 0.97 & 0.03 \\ 
  2015/01 & 10 & 0.82 & 0.18 & 5 & 0.93 & 0.07 & 3 & 0.85 & 0.15 & 2 & 0.98 & 0.02 & 3 & 0.85 & 0.15 & 2 & 0.98 & 0.02 \\ 
  2015/02 & 5 & 0.86 & 0.14 & 5 & 0.90 & 0.10 & 5 & 0.86 & 0.14 & 3 & 0.91 & 0.09 & 5 & 0.86 & 0.14 & 6 & 0.90 & 0.10 \\ 
  2015/03 & 5 & 0.90 & 0.10 & 5 & 0.96 & 0.04 & 5 & 0.90 & 0.10 & 3 & 0.96 & 0.04 & 5 & 0.90 & 0.10 & 6 & 0.96 & 0.04 \\ 
  2015/04 & 5 & 0.90 & 0.10 & 5 & 0.94 & 0.06 & 5 & 0.90 & 0.10 & 3 & 0.95 & 0.05 & 5 & 0.90 & 0.10 & 6 & 0.94 & 0.06 \\ 
  2015/05 & 10 & 0.90 & 0.10 & 15 & 0.96 & 0.04 & 3 & 0.92 & 0.08 & 3 & 0.96 & 0.04 & 3 & 0.92 & 0.08 & 3 & 0.96 & 0.04 \\ 
  2015/06 & 10 & 0.87 & 0.13 & 15 & 0.86 & 0.14 & 3 & 0.90 & 0.10 & 3 & 0.88 & 0.12 & 3 & 0.90 & 0.10 & 3 & 0.88 & 0.12 \\ 
  2015/07 & 10 & 0.88 & 0.12 & 15 & 0.48 & 0.52 & 3 & 0.90 & 0.10 & 3 & 0.48 & 0.52 & 3 & 0.90 & 0.10 & 3 & 0.48 & 0.52 \\ 
  2015/08 & 10 & 0.88 & 0.12 & 5 & 0.59 & 0.41 & 2 & 0.93 & 0.07 & 1 & 1.00 & 0.00 & 12 & 0.88 & 0.12 & 93 & 0.58 & 0.42 \\ 
  2015/09 & 10 & 0.89 & 0.11 & 5 & 0.53 & 0.47 & 2 & 0.93 & 0.07 & 1 & 1.00 & 0.00 & 12 & 0.88 & 0.12 & 93 & 0.52 & 0.48 \\ 
  2015/10 & 10 & 0.90 & 0.10 & 5 & 0.59 & 0.41 & 2 & 0.96 & 0.04 & 1 & 1.00 & 0.00 & 12 & 0.90 & 0.10 & 93 & 0.58 & 0.42 \\ 
  2015/11 & 10 & 0.92 & 0.08 & 5 & 0.82 & 0.18 & 2 & 0.97 & 0.03 & 2 & 0.83 & 0.17 & 2 & 0.97 & 0.03 & 9 & 0.82 & 0.18 \\ 
  2015/12 & 10 & 0.93 & 0.07 & 5 & 0.84 & 0.16 & 2 & 0.98 & 0.02 & 2 & 0.84 & 0.16 & 2 & 0.98 & 0.02 & 9 & 0.84 & 0.16 \\ 
  2016/01 & 10 & 0.92 & 0.08 & 5 & 0.87 & 0.13 & 2 & 0.97 & 0.03 & 2 & 0.87 & 0.13 & 2 & 0.97 & 0.03 & 9 & 0.87 & 0.13 \\ 
  2016/02 & 5 & 0.89 & 0.11 & 10 & 0.90 & 0.10 & 1 & 1.00 & 0.00 & 1 & 1.00 & 0.00 & 16 & 0.87 & 0.13 & 1 & 1.00 & 0.00 \\ 
  2016/03 & 5 & 0.83 & 0.17 & 10 & 0.86 & 0.14 & 1 & 1.00 & 0.00 & 1 & 1.00 & 0.00 & 16 & 0.81 & 0.19 & 1 & 1.00 & 0.00 \\ 
  2016/04 & 5 & 0.85 & 0.15 & 10 & 0.93 & 0.07 & 1 & 1.00 & 0.00 & 1 & 1.00 & 0.00 & 16 & 0.84 & 0.16 & 1 & 1.00 & 0.00 \\ 
  2016/05 & 15 & 0.83 & 0.17 & 10 & 0.75 & 0.25 & 3 & 0.87 & 0.13 & 1 & 1.00 & 0.00 & 3 & 0.87 & 0.13 & 3 & 0.75 & 0.25 \\ 
  2016/06 & 15 & 0.85 & 0.15 & 10 & 0.65 & 0.35 & 3 & 0.88 & 0.12 & 1 & 1.00 & 0.00 & 3 & 0.88 & 0.12 & 3 & 0.65 & 0.35 \\ 
  2016/07 & 15 & 0.84 & 0.16 & 10 & 0.71 & 0.29 & 3 & 0.90 & 0.10 & 1 & 1.00 & 0.00 & 3 & 0.90 & 0.10 & 3 & 0.71 & 0.29 \\ 
  2016/08 & 15 & 0.83 & 0.17 & 5 & 0.70 & 0.30 & 3 & 0.90 & 0.10 & 2 & 0.72 & 0.28 & 3 & 0.90 & 0.10 & 2 & 0.72 & 0.28 \\ 
  2016/09 & 15 & 0.82 & 0.18 & 5 & 0.73 & 0.27 & 3 & 0.88 & 0.12 & 2 & 0.77 & 0.23 & 3 & 0.88 & 0.12 & 2 & 0.77 & 0.23 \\ 
  2016/10 & 15 & 0.84 & 0.16 & 5 & 0.84 & 0.16 & 3 & 0.89 & 0.11 & 2 & 0.85 & 0.15 & 3 & 0.89 & 0.11 & 2 & 0.85 & 0.15 \\ 
  2016/11 & 15 & 0.87 & 0.13 & 10 & 0.94 & 0.06 & 3 & 0.91 & 0.09 & 1 & 1.00 & 0.00 & 3 & 0.91 & 0.09 & 1 & 1.00 & 0.00 \\ 
  2016/12 & 15 & 0.89 & 0.11 & 10 & 0.94 & 0.06 & 3 & 0.93 & 0.07 & 1 & 1.00 & 0.00 & 3 & 0.93 & 0.07 & 1 & 1.00 & 0.00 \\ 
  2017/01 & 15 & 0.88 & 0.12 & 10 & 0.92 & 0.08 & 3 & 0.93 & 0.07 & 1 & 1.00 & 0.00 & 3 & 0.93 & 0.07 & 1 & 1.00 & 0.00 \\ 
  2017/02 & 10 & 0.89 & 0.11 & 5 & 0.93 & 0.07 & 2 & 0.94 & 0.06 & 2 & 0.94 & 0.06 & 4 & 0.92 & 0.08 & 2 & 0.94 & 0.06 \\ 
  2017/03 & 10 & 0.81 & 0.19 & 5 & 0.74 & 0.26 & 2 & 0.87 & 0.13 & 2 & 0.82 & 0.18 & 4 & 0.84 & 0.16 & 2 & 0.82 & 0.18 \\ 
   \hline
\hline
\end{tabular}
\endgroup
\caption{Average monthly weights of BTC and altcoins in the respective periods in the 6 indices} 
\label{table_weights}
\end{sidewaystable}

\end{document}